\documentclass
[prc,superscriptaddress,twocolumn,showpacs,preprintnumbers,amsmath,amssymb,dvipdfmx]{revtex4-1}
\usepackage{graphicx}
\usepackage{dcolumn}
\usepackage{amsmath}
\usepackage{times}
\usepackage{braket}
\usepackage{multirow}
\def\fun#1#2{\lower3.6pt\vbox{\baselineskip0pt\lineskip.9pt
 \ialign{$\mathsurround=0pt#1\hfil##\hfil$\crcr#2\crcr\sim\crcr}}}
\usepackage{graphicx}% Include figure files
\usepackage{dcolumn}% Align table columns on decimal point
\usepackage{bm}% bold math
\newcommand{\vect}{\boldsymbol}
\usepackage{ulem}
\usepackage{color}
\definecolor{verde}{rgb}{0.0,0.7,0.0}

\begin{document}

%\preprint{APS/123-QED}

\title{Theoretical investigation of two-particle two-hole effect \\on spin-isospin excitations through charge-exchange reactions}


\author{Tokuro Fukui}
\email{fukui@na.infn.it}
\affiliation{Istituto Nazionale di Fisica Nucleare, Complesso Universitario di Monte S. Angelo, Via Cintia, I-80126 Napoli, Italy}
\affiliation{Nuclear Data Center, Japan Atomic Energy Agency, Tokai, Ibaraki 319-1195, Japan}

\author{Futoshi Minato}
\email{minato.futoshi@jaea.go.jp}
\affiliation{Nuclear Data Center, Japan Atomic Energy Agency, Tokai, Ibaraki 319-1195, Japan}
\affiliation{NSCL/FRIB Laboratory, Michigan State University, East Lansing, Michigan 48824, USA}

\date{\today}% It is always \today, today,

\begin{abstract}
 \noindent
 \textbf{Background:}
 Coherent one-particle one-hole (1p1h) excitations have given us effective insights into general nuclear excitations.
 However, the two-particle two-hole (2p2h) excitation beyond 1p1h is now recognized as critical
 for the proper description of experimental data of various nuclear responses.\\
 \textbf{Purpose:}
 The spin-flip charge-exchange reactions $^{48}{\rm Ca}(p,n)^{48}{\rm Sc}$ are investigated to clarify the role of the 2p2h effect on their cross sections.
 The Fermi transition of $^{48}{\rm Ca}$ via the $(p,n)$ reaction is also investigated in order to demonstrate our framework. \\
 \textbf{Methods:}
 The transition density is calculated microscopically with the second Tamm-Dancoff approximation, and the distorted-wave Born approximation
 is employed to describe the reaction process. A phenomenological one-range Gaussian interaction is used to prepare the form factor.\\
 \textbf{Results:}
 For the Fermi transition, our approach describes the experimental behavior of the cross section better than the Lane model,
 which is the conventional method.
 For spin-flip excitations including the GT transition, the 2p2h effect decreases the magnitude of the cross section
 and does not change the shape of the angular distribution.
 The $\Delta l=2$ transition of the present reaction is found to play a negligible role. \\
\textbf{Conclusions:}
 The 2p2h effect will not change the angular-distributed cross section of spin-flip responses.
 This is because the transition density of the Gamow-Teller response, the leading contribution to the cross section,
 is not significantly varied by the 2p2h effect.
\end{abstract}

\pacs{}% PACS, the Physics and Astronomy
\maketitle

%\tableofcontents

%%%%%%%%%%%%%%%%%%%%%%%%%%%%%%%%%%%%%%%%%%%%%%%%%%%%%%%%%%%%%%%%%
\section{Introduction}
\label{Intro}
%%%%%%%%%%%%%%%%%%%%%%%%%%%%%%%%%%%%%%%%%%%%%%%%%%%%%%%%%%%%%%%%%
Several theories based on the picture of particle-hole excitations have successfully described the nuclear spin-isospin response.
Not only the lowest-order one-particle one-hole (1p1h) excitation but also many-particle many-hole
excitations have been observed to play a significant role. For neutrino-nucleus quasielastic scattering,
the inclusion of the two-particle two-hole (2p2h) configuration enhances the cross section to the point where a theoretical calculation is
consistent with the experimental data~\cite{MartiniPhysRevC.80.065501}.
For a nuclear spin-isospin response, the 2p2h-configuration mixing partly explains the Gamow-Teller (GT) quenching problem,
which is known not to follow the Ikeda sum rule~\cite{Ikeda01031964,FUJITA1965145,Gaarde1983}
and is one of the long-standing problems in nuclear physics.
Other reasons for this quenching are the existence of the $\Delta$-$h$ excitation
and the coupling of the GT-$1^+$ state to the spin-quadrupole (SQ) $1^+$ state mediated by the tensor force \cite{BAI2009}.
For a more detailed review, see Ref.~\cite{Prog.Part.Nucl.Phys.56.446}.

Recently the effect of 2p2h mixing on the GT-transition strength $B({\rm GT})$ employing the fully self-consistent second Tamm-Dancoff approximation (STDA)
was investigated~\cite{MinatoPhysRevC.93.044319}.
The calculated $B({\rm GT})$ distribution of $^{48}{\rm Ca}$ was compared with its experimentally measured value~\cite{YakoPhysRevLett.103.012503},
which was derived through an analysis of the charge-exchange reaction
$^{48}{\rm Ca}(p,n)^{48}{\rm Sc}$ by means of the distorted-wave impulse approximation (DWIA).
It was shown that the STDA (where the 2p2h model space was confined to seven single-particle levels) described the experimental data
better than the 1p1h Tamm-Dancoff approximation (TDA).
It was also confirmed that the broadening of the $B({\rm GT})$ distribution by the 2p2h effect was essential
to account for the observed experimental behavior, as preceding works have shown using different models~\cite{Phys.Lett.166B.18,Phys.Rev.C90.054328}.

Charge exchange reactions such as  $(p,n)$ and $(^3{\rm He},t)$ excite target nuclei
through the one-body operator which populates 1p1h states in target nuclei.
Therefore, the 2p2h configuration is not directly involved in the above reactions.
However, it would have an indirect influence on the cross sections through the transition from 1p1h states to 2p2h states.
Even though the importance of the 2p2h configuration
in $B({\rm GT})$ has been pointed out by many authors~\cite{Phys.Lett.166B.18, Wakasa2012, Prog.Part.Nucl.Phys.56.446, Wambach1988},
a microscopic understanding of how such a higher-order configuration
is involved in the charge-exchange cross section is not so transparent.
In the present paper we therefore investigate the effect of the 2p2h configuration on the angular-distributed cross section,
which is directly comparable with experimental data.
In order to calculate the cross section of spin-flip transitions, we work with the distorted-wave Born approximation (DWBA)
with the microscopic transition density obtained by the STDA.

The $\Delta l=2$ transition also generates $1^+$ resonance states
at excitation energies equal to that of GT-$1^+$ states.
When the experimental $B({\rm GT})$ of $^{48}{\rm Ca}$ was evaluated~\cite{YakoPhysRevLett.103.012503},
it was assumed that the zero-degree charge-exchange cross section is proportional to $B({\rm GT})$
and the $\Delta l=2$ transition plays a negligible role.
The assumption, however, was confirmed only for the $^{13}{\rm C}(p,n)^{13}{\rm N}$ reaction~\cite{Taddeucci1987}.
Therefore we investigate whether the assumption is valid also for the present system, $^{48}{\rm Ca}(p,n)^{48}{\rm Sc}$.

In this article both the spin-flip and non-spin-flip transitions are surveyed.
Note that, in the Fermi transition, the nucleus populates the isobaric analogue state (IAS),
which has the same isospin $T_A$ as that of the ground state of the parent nucleus.
It does not couple to other 1p1h states having $T_A \pm 1$;
i.e., the 2p2h effect is expected to be small for the Fermi transition as seen in outgoing neutron spectra of $(p,n)$ reactions
(see, for example, Ref.~\cite{PhysRevC.31.1147}).
Therefore we investigate the Fermi transition just to demonstrate our theoretical framework.

% This work is also motivated from the engineering point of view, especially in the field of accelerators.
% For example, neutron generated by $(p,n)$ reactions activates materials used in accelerators.
% For the purpose of radiation protection and decommissioning of the accelerator,
% it is important to know nuclear data of $(p,n)$ reactions as precisely as possible.

This paper is organized as follows. Section~\ref{formulation} is dedicated to the formulation
of the structure and reaction models as well as the form factor.
In Sec.~\ref{result}, results on the non-spin-flip (Fermi-type) transition are shown.
The 2p2h effect on the cross section of the spin-flip transition is also discussed.
A summary is given in Sec.~\ref{summary}.

%%%%%%%%%%%%%%%%%%%%%%%%%%%%%%%%%%%%%%%%%%%%%%%%%%%%%%%%%%%%%%%%%
\section{Theoretical framework}
\label{formulation}
Convolution of the DWBA and the random phase approximation (which includes a correlation in ground states in addition to the TDA)
is widely used to analyze experimental data at intermediate energies, as in the measurements of charge-exchange cross sections
at zero degree~\cite{Taddeucci1987,Phys.Rev.Lett.59.1401,Phys.Rev.C91.064316}
accompanied by the multipole decomposition technique~\cite{Prog.Part.Nucl.Phys.56.446}.
Here we use DWBA+TDA/STDA based on the Skyrme energy density functional~\cite{VautherinBrink}
in order to discuss the effect of the 2p2h configuration on charge-exchange cross sections.
Because the ground state correlation is expected to be small for the Fermi- and GT-type charge-exchange reactions~\cite{NISHIZAKI1988231,NGUYENDINHDANG1997719}, the use of the TDA will not affect our results.
We briefly describe the STDA as well as the TDA in Sec.~\ref{formulation1} and illustrate the formulations of the form factor and the DWBA
in Secs.~\ref{formulation2} and~\ref{formulation3}, respectively.
%%%%%%%%%%%%%%%%%%%%%%%%%%%%%%%%%%%%%%%%%%%%%%%%%%%%%%%%%%%%%%%%%
%%%%%%%%%%%%%%%%%%%%%%%%%%%%%%%%%%%%%%%%%%%%%%%%%%%%%%%%%%%%%%%%%
\subsection{Structure model}
\label{formulation1}
%%%%%%%%%%%%%%%%%%%%%%%%%%%%%%%%%%%%%%%%%%%%%%%%%%%%%%%%%%%%%%%%%
We consider the transition $A \rightarrow B$ induced by the charge-exchange reaction $A(p,n)B$.
To describe the GT and $\Delta l=2$ transitions as well as other $1^+$ multipole spin-flip transitions relevant to the reaction studied,
we adopt the STDA as explained in Ref.~\cite{MinatoPhysRevC.93.044319}.
In the STDA the many-body wave function $\ket{B_{\alpha}}$, which is a resonance state of $B$ with respect to the $A$'s ground state
$\ket{A}$, is written as
\begin{align}
 \ket{B_\alpha}
 &=
 \left[
 \sum_{mi}X_{mi} a_m^\dagger a_i
 +\sum_{mnij}\mathcal{X}_{mnij} a_m^\dagger a_n^\dagger a_i a_j
 \right]
 \ket{A}
 \nonumber\\
 &\equiv
 \sum_{mi}X_{mi} \ket{m(i)^{-1}}
 +
 \sum_{mnij}\mathcal{X}_{mnij} \ket{mn(ij)^{-1}},
 \label{STDAOpe}
\end{align}
where $a_\nu^\dagger$ ($a_\nu$) is the creation (annihilation) operator in the single-particle state $\nu$,
and $\nu=m,n,p,q$ ($\nu=i,j,k,l$) for the particle (hole) states. We introduce the index $\alpha$ to express the non-spin-flip transition ($\alpha=s0$),
the spin-flip transition ($\alpha=s1$), and the $\Delta l=2$ transition ($\alpha=l2$).
We work with the Skyrme-Hartree-Fock model to obtain $\ket{A}$.
The coefficients $X_{mi}$ and $\mathcal{X}_{mnij}$ are determined by solving the so-called STDA equation~\cite{Yannouleas1987},
\begin{align}
\left(
 \begin{array}{cc}
  A & \mathcal{A}_{12} \\
  \mathcal{A}_{21} & \mathcal{A}_{22} \\
 \end{array}
 \right)
\left(
 \begin{array}{c}
  X \\
 \mathcal{X}\\
 \end{array}
 \right)
=
\varepsilon
\left(
 \begin{array}{c}
  X \\
 \mathcal{X}\\
 \end{array}
 \right).
 \label{STDAeq}
\end{align}
Here the matrix elements in Eq.~\eqref{STDAeq} are given by Ref.~\cite{Yannouleas1987} and $\varepsilon$ is the phonon energy.

A value of $\mathcal{X}_{mnij}=0$ corresponds to the standard TDA,
which does not include 2p2h-configuration mixing.
In the TDA, $\Ket{B_\alpha}$ is given by
\begin{align}
 \Ket{B_\alpha}
 =
 \sum_{mi}
 X_{mi} a_m^\dagger a_i
 \ket{A}.
 \label{TDAOpe}
\end{align}
The coefficients $X_{mi}$ are obtained from the so-called TDA equation; see Refs.~\cite{DJRowe,RingandSchuck}.
Since the 2p2h effect on the IAS originating from the Fermi transition is known to be negligible (see Sec.~\ref{Intro}),
we describe $\Ket{B_{s0}}$ by Eq.~\eqref{TDAOpe}.

The transition density, which is employed to calculate the form factor shown later, is given by
\begin{align}
 g_\alpha(r_{i_t})
 &=
 \frac{1}{\hat{j}_B} \sum_{mi}
 X_{mi}R_m(r_{i_t})R_i(r_{i_t})
 \left\langle j_ml_m\left|\left|
 \mathcal{G}_\alpha
 \right|\right|j_il_i\right\rangle,
 \label{traden}
\end{align}
where $R_m$ $(R_i)$ is the radial part of the single-particle wave-function
and $j_m = l_m \pm 1/2$ ($j_i = l_i \pm 1/2$) with the magnitude of the orbital angular momentum $l_m$ ($l_i$) of state $m$ $(i)$.
The coordinate of the $i_t$ th nucleon in the target is $\vect{r}_{i_t}$ and $j_B$ is the magnitude of the spin of $B$.
We use the abbreviation $\hat{j}_B=\sqrt{2j_B+1}$.
The transition operator for the non-spin-flip transition is
\begin{align}
 \mathcal{G}_{s0}=\vect{\tau}Y_{l=0,0}(\hat{\vect{r}}_{i_t}),
 \label{GF}
\end{align}
and those of the spin-flip and $\Delta l=2$ transitions are, respectively,
\begin{align}
 \mathcal{G}_{s1}&=\vect{\tau}Y_{l=0,0}(\hat{\vect{r}}_{i_t})\boldsymbol{\sigma}, %_{1M},
 \label{GGT}\\
 \mathcal{G}_{l2}&=\vect{\tau}\left[Y_{l=2}(\hat{\vect{r}}_{i_t})\otimes\boldsymbol{\sigma}\right]_{1M},
 \label{Gl2}
\end{align}
where $\vect{\sigma}$ ($\vect{\tau}$) is the Pauli spin (isospin) operator.
Here $l$ corresponds to the orbital angular momentum transfer of the relative motion [see Eq.~\eqref{transAM}], and $M=0,\pm 1$.

%%%%%%%%%%%%%%%%%%%%%%%%%%%%%%%%%%%%%%%%%%%%%%%%%%%%%%%%%%%%%%%%%
\subsection{Form factor}
\label{formulation2}
%%%%%%%%%%%%%%%%%%%%%%%%%%%%%%%%%%%%%%%%%%%%%%%%%%%%%%%%%%%%%%%%%
The form factor is expressed by
\begin{align}
 \mathcal{F}_{\alpha}(\vect{R})
 =\Braket{nB\left|v_{\alpha}\right|pA},
 \label{FF1}
\end{align}
where $\vect{R}$ is the relative coordinate of the $p$-$A$ and $n$-$B$ system.
The ket (bra) vector represents the product of the spin-wave function of the projectile (ejectile) and the many-body wave function of $A$ ($B$).
The transitions of non-spin-flip (the spin transfer $\Delta s=0$), spin-flip ($\Delta s=1$), and $\Delta l=2$ components
are respectively caused by the interactions,
\begin{align}
 v_{s0}&=\sum_{i_pi_t}V_{s0}(\rho)\vect{\tau}_{i_p}\cdot\vect{\tau}_{i_t},
 \label{vF}\\
 v_{s1}&=\sum_{i_pi_t}V_{s1}(\rho)\left(\vect{\sigma}_{i_p}\cdot\vect{\sigma}_{i_t}\right)
 \left(\vect{\tau}_{i_p}\cdot\vect{\tau}_{i_t}\right),
 \label{vGT}\\
 v_{l2}&=\sum_{i_pi_t}V_{l2}(\rho)\left([\vect{\sigma}_{i_p}Y_2]_1\cdot[\vect{\sigma}_{i_t}Y_2]_1\right)
 \left(\vect{\tau}_{i_p}\cdot\vect{\tau}_{i_t}\right),
 \label{vl2}
\end{align}
where $\rho=\left|r_{i_p}-r_{i_t}\right|$ %$\vect{\rho}=\vect{r}_{i_p}-\vect{r}_{i_t}$
and the sums over $i_p$ ($i_t$) represent the nucleon number of the projectile (target nucleus) running up to its mass number.

We assume that the radial parts $V_{s0}$, $V_{s1}$, and $V_{l2}$ are the one-range Gaussian functions given by
\begin{align}
 V_{s0}(\rho)=\bar{V}_{0}e^{-\left(\frac{\rho}{\rho_0}\right)^2},
 \quad
 V_{s1}(\rho)=V_{l2}(\rho)=\bar{V}_{1}e^{-\left(\frac{\rho}{\rho_1}\right)^2},
\end{align}
the parameters of which are determined phenomenologically.
Since the present work focuses on the investigation of the 2p2h effect,
we use this phenomenological interaction rather than a microscopic one.

Following the formalism of Refs.~\cite{Petrovich1977,Cook1984}, the form factor is obtained through the partial-wave expansion.
The radial part of $\mathcal{F}_{\alpha}$ is calculated as
\begin{align}
 F_{lsj}^\alpha(R)=\frac{i^l}{\pi^2}\hat{j}\int \tilde{V}_\alpha (K) \tilde{g}_{\alpha} (K)j_l(KR)K^2 dK ,
 \label{radFF}
\end{align}
where $s$ and $j$ are the transferred angular momenta defined by
\begin{align}
 \vect{j}=\vect{j}_B-\vect{j}_A,\quad \vect{s}=\vect{j}_p-\vect{j}_n,\quad \vect{l}=\vect{j}-\vect{s},
 \label{transAM}
\end{align}
with the spin $\vect{j}_x$ of the particle $x$~$(=p,n,A,~{\rm and}~B)$.
The interaction and transition density in momentum space, regarding $K$ associated with $R$, are respectively defined by
\begin{align}
 \tilde V_\alpha (K)
 &=
 4\pi \int d\rho \rho^2 \frac{\sin(K\rho)}{K\rho} V_\alpha(\rho),
 \label{momv}\\
 \tilde g_{\alpha} (K)
 &=
 4\pi \int dr_{i_t} \, r_{i_t}^2 j_{l} (Kr_{i_t}) g_\alpha(r_{i_t})
 \label{momg},
\end{align}
with the spherical Bessel function $j_l$ ($l=0$ for the non-spin-flip and spin-flip transitions, and $l=2$ for the $\Delta l=2$ case).

Expanding the spherical Bessel function of Eq.~\eqref{momg} in terms of $K$, we obtain the integrands proportional to
$g_\alpha, r_{i_t}^2g_\alpha$, and so on.
The lowest-order terms for $\alpha=s1$ and $l2$ correspond to the GT- and SQ-$1^+$ transitions, respectively.
Incidentally, the first order for $\alpha=s1$ is the spin-monopole transition,
which is difficult to distinguish from the GT transition experimentally~\cite{YakoPhysRevLett.103.012503}.
Higher-order contributions can be safely ignored, as they are negligibly small in excitation energies studied in this work.

%%%%%%%%%%%%%%%%%%%%%%%%%%%%%%%%%%%%%%%%%%%%%%%%%%%%%%%%%%%%%%%%%
\subsection{Reaction model}
\label{formulation3}
%%%%%%%%%%%%%%%%%%%%%%%%%%%%%%%%%%%%%%%%%%%%%%%%%%%%%%%%%%%%%%%%%
The following expression is based on the formalism in Ref.~\cite{Cook1984}
but now generalized to include the spin-orbit interaction regarding the coupling between
the projectile's (ejectile's) spin and the $p$-$A$ ($n$-$B$) orbital angular momentum in the initial (final) channel.
The transition matrix with the DWBA under the partial-wave expansion is given by
\begin{align}
 T^{\rm (DWBA)}_{\alpha;m_p m_n m_A m_B}
 &=\frac{4\pi}{K_pK_n}(-)^{j_n+m_n} \hat{j}_n
 \nonumber\\
 &\times
 \sum_{jm_j}\left( j_A m_A j m_j | j_B m_B \right)
 \mathcal{S}_{\alpha;jm_j}^{m_pm_n},
 \label{TmatGen}
\end{align}
where $m_j$ and $m_x$ respectively correspond to the $z$-projections of $\vect{j}$ and $\vect{j}_x$.
The magnitude of the wave number of the projectile (ejectile) is expressed by $K_p$ ($K_n$).
The function $\mathcal{S}_{\alpha;jm_j}^{m_pm_n}$ is defined by
\begin{align}
 \mathcal{S}_{\alpha;jm_j}^{m_pm_n}
 &=
 \left(4\pi\right)^{-\frac{1}{2}}
 \sum_{\substack{J_iJ_f\\L_iL_f\\ls}}
 i^{L_i-L_f-l} \hat{s}\hat{J}_i\hat{J}_f\hat{L}_i^2\hat{L}_f^2
 I_{J_iJ_fL_iL_f}^{\alpha;lsj}
 \nonumber\\
 &\times
 \left( L_i 0 L_f 0 | l 0 \right)
 \begin{Bmatrix}
  L_f & L_i & l \\
  j_n & j_p & s \\
  J_f & J_i & j
 \end{Bmatrix}
 f_{jj_pj_nL_iL_f}^{m_jm_pm_n} \left(\cos\theta\right),
 \label{redTmatGen}
\end{align}
where $L_i$ ($L_f$) is the magnitude of the orbital angular momentum regarding the relative $p$-$A$ ($n$-$B$) motion,
and its coupled spin with $j_p$ ($j_n$) is expressed by $J_i$ ($J_f$).
The conservation of the total angular momentum is given by
\begin{align}
 \left[\left[\vect{j}_p \otimes \vect{L}_i\right]_{\vect{J}_i} \otimes \vect{j}_A\right]
 =
 \left[\left[\vect{j}_n \otimes \vect{L}_f\right]_{\vect{J}_f} \otimes \vect{j}_B\right].
 \label{totalJ}
\end{align}
The overlap integral $I_{J_iJ_fL_iL_f}^{\alpha;lsj}$ and the function $f_{jj_pj_nL_iL_f}^{m_jm_pm_n}$ are respectively defined as
\begin{align}
 I_{J_iJ_fL_iL_f}^{\alpha;lsj}&=\int dR \tilde \xi_{n;J_fL_f}(K_f, R)F_{lsj}^\alpha(R)\tilde \xi_{p;J_iL_i}(K_i, R),
 \label{ovlI}
\end{align}
\begin{align}
 &f_{jj_pj_nL_iL_f}^{m_jm_pm_n} \left(\cos\theta\right)
 \nonumber\\
 &\quad=
 \left( J_i m_p J_f, m_j-m_p | j m_j \right)
 \left( j_p m_p L_i 0 | J_i m_p \right)
 \nonumber\\
 &\quad\times
 \left( j_n, -m_n L_f, m_j-m_p+m_n | J_f, m_j-m_p \right)
 \nonumber\\
 &\quad\times
 \left[ \frac{(L_f-\left|m_j-m_p+m_n \right|)!}{(L_f+\left|m_j-m_p+m_n\right|)!} \right]^{\frac{1}{2}}
 P_{L_f,m_j-m_p+m_n} \left(\cos\theta\right),
 \label{fmmm}
\end{align}
with the Legendre function $P_{L_f,m_j-m_p+m_n}$ as a function of the emitting angle $\theta$.

The partial wave $\tilde\xi_{\gamma;JL}~=~P_{\rm NL}^{(\gamma)}\xi_{\gamma;JL}$ ($\gamma=p~{\rm or}~n$)
is given as the solution of the Schr\"odinger equation,
\begin{align}
 \left[\frac{d^2}{dR^2} +K_\gamma -\frac{L(L+1)}{R^2} -\frac{2\mu_\gamma}{\hbar^2}U_\gamma(R) \right]
 \xi_{\gamma;JL}(K_\gamma,R) =0,
 \label{ScheqLS}
\end{align}
where the reduced mass is represented by $\mu_\gamma$ and
the distorting potential $U_\gamma$ involves the central, spin-orbit, and Coulomb terms.
Here, in order to take into account the nonlocality of the nucleon optical potential,
we multiply the distorted wave $\xi_{\gamma;JL}$ by the so-called Perey factor $P_{\rm NL}^{(\gamma)}$~\cite{PEREY1962353},
\begin{align}
 P_{\rm NL}^{(\gamma)}(R)
 &=
 \left[
 1-\frac{\mu_p\beta^2}{2\hbar^2}U_\gamma^{\rm (N)}(R)
 \right]^{-\frac{1}{2}},
 \label{Pereyfac}
\end{align}
with the nonlocal parameter $\beta$ and the nuclear part $U_\gamma^{\rm (N)}$ of the distorting potential.

The cross section is calculated as
\begin{align}
 \frac{d\sigma_\alpha}{d\Omega}
 &=
 \frac{\mu_p\mu_n}{\left(2\pi\hbar^2\right)^2}\frac{K_n}{K_p}\frac{1}{\left(\hat{j}_p\hat{j}_A\right)^2}
 \sum_{\substack{m_pm_n\\m_Am_B}}\left|T^{\rm (DWBA)}_{\alpha;m_p m_n m_A m_B}\right|^2
 \nonumber\\
 &=
 \frac{1}{E_pE_n}\frac{K_n}{K_p}\left(\frac{\hat{j}_n\hat{j}_B}{\hat{j}_p\hat{j}_A}\right)^2
 \sum_{jm_j}\frac{1}{\hat{j}^2}\sum_{m_pm_n}\left|\mathcal{S}_{\alpha;jm_j}^{m_pm_n}\right|^2
 \label{csGen},
\end{align}
with $E_\gamma=\left(\hbar K_\gamma\right)^2/(2\mu_\gamma)$.

%%%%%%%%%%%%%%%%%%%%%%%%%%%%%%%%%%%%%%%%%%%%%%%%%%%%%%%%%%%%%%%%%
\section{Results and discussion}
\label{result}
%%%%%%%%%%%%%%%%%%%%%%%%%%%%%%%%%%%%%%%%%%%%%%%%%%%%%%%%%%%%%%%%%
%%%%%%%%%%%%%%%%%%%%%%%%%%%%%%%%%%%%%%%%%%%%%%%%%%%%%%%%%%%%%%%%%
\subsection{Model setting}
\label{result1}
%%%%%%%%%%%%%%%%%%%%%%%%%%%%%%%%%%%%%%%%%%%%%%%%%%%%%%%%%%%%%%%%%
The ground state wave function of $^{48}{\rm Ca}$ is calculated by the Skyrme-Hartree-Fock approach~\cite{VautherinBrink}
with the SGII effective interaction \cite{Phys.Lett.B106_379}.
To obtain the non-spin-flip $0^+$ and spin-flip $1^+$ excited states, we solve the STDA and TDA equations with the same force in a self-consistent manner,
and the transition density given by Eq.~\eqref{traden} is calculated for each state.
The model space of the STDA and TDA calculations consists of single-particle orbits up to $100$~MeV for the 1p1h configuration
and $1d_{5/2}, 1d_{3/2}, 2s_{1/2}, 1f_{7/2}, 2p_{3/2}, 2p_{1/2}$, and $1f_{5/2}$ orbits for the 2p2h configuration
as performed in Ref.~\cite{MinatoPhysRevC.93.044319}.
The neutron and proton orbits are assumed to be fully occupied up to $1f_{7/2}$ and $2s_{1/2}$, respectively.

To calculate the form factor, we adjust the strengths $\bar{V}_0$ and $\bar{V}_1$,
while keeping the range parameter fixed at $\rho_0=\rho_1=1.484$~fm~\cite{Ohmura1970}.
% For the non-spin-flip transition, we let $\bar{V}_0=-712.1$~MeV in order to fit the calculated cross section to the measured data at forward angle.
For the non-spin-flip transition, we let $\bar V_0=-1762.4$~MeV in order to fit the calculated cross section to the measured data at forward angle.
For the spin-flip transitions, we use $\bar{V}_1=-275.8$~MeV and $-153.9$~MeV for the low-lying and giant resonances respectively,
to make the calculated cross section with the STDA transition density identical to the measured data at 0.2$^\circ$.
The same parameters $\bar{V}_1$ and $\rho_1$ are used in the calculation of the form factor with the TDA transition density.
%\bar{V}_0=72.15*pi^2=712.1
%\bar{V}_1(low-lying)=48.40*pi^2/sqrt(3)=275.8
%\bar{V}_1(Giant)=27.00*pi^2/sqrt(3)=153.9

For $U_\gamma^{\rm (N)}$, we adopt the phenomenological optical potential~\cite{KONING2003231}
and the ``Fit 1'' parameter set of the Dirac phenomenology~\cite{HamaPhysRevC.41.2737},
for the non-spin-flip and the spin-flip transitions, respectively.
We also include the prescription~\cite{SatchlerPhysRev.136.B637} that the incident energy dependence of the optical potential
for the Fermi transition should be adjusted as $E_{\rm lab}-Q/2$, where $E_{\rm lab}$ is the incident energy in the laboratory frame
and $Q$ stands for the $Q$ value.
The nonlocal range parameter is $\beta=0.85$~fm~\cite{PEREY1962353}, and the Coulomb potential is chosen to be
a uniformly charged sphere with the charge radius of 4.61~fm~\cite{KONING2003231}.
The partial wave $\xi_{\gamma}$ is calculated up to $J=20.5$ ($J=100.5$) for the non-spin-flip (spin-flip) transition.
For each transition the integration in Eq.~\eqref{ovlI} is performed up to 20~fm; our work assumes the relativistic kinematics.

%%%%%%%%%%%%%%%%%%%%%%%%%%%%%%%%%%%%%%%%%%%%%%%%%%%%%%%%%%%%%%%%%
\subsection{Non-spin-flip transitions}
\label{result2}
%%%%%%%%%%%%%%%%%%%%%%%%%%%%%%%%%%%%%%%%%%%%%%%%%%%%%%%%%%%%%%%%%
%%%%%%%%%%%%%%%
%%% Figure  %%%
%%%%%%%%%%%%%%%
\begin{figure}[!b]
\begin{center}
\includegraphics[width=0.48\textwidth,clip]{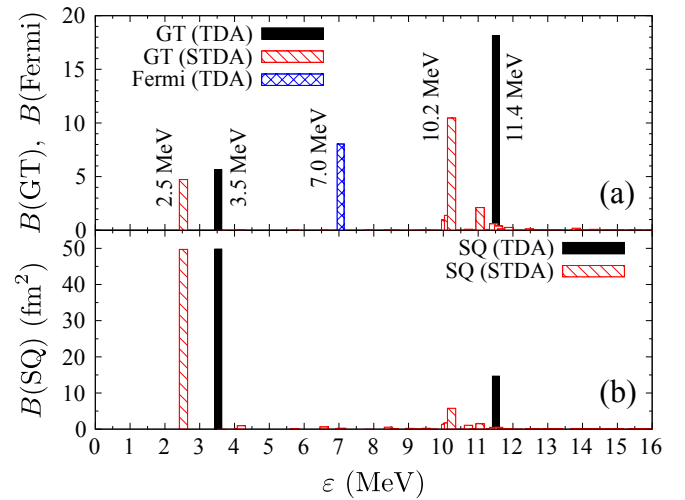}
 \caption{(a) Strength functions of the GT and IAS resonances of $^{48}{\rm Ca}$ calculated with the STDA and TDA.
 The filled and slash-shaded bars are the results of the GT transition calculated with the TDA and STDA, respectively, and the cross-shaded bars are for the IAS resonance.
 (b) Strength functions of the SQ-1$^+$ transition.}
\label{strength}
\end{center}
\end{figure}

To demonstrate our model, we first discuss the Fermi transition measured from the $^{48}{\rm Ca}(p,n)^{48}{\rm Sc(IAS)}$ reaction.
Figure \ref{strength} shows the strength functions of the Fermi and GT transitions of $^{48}{\rm Ca}$ calculated with the STDA and TDA.
The corresponding excitation energies of the resonance states in question are written explicitly in Fig.~\ref{strength}(a).
The TDA calculation gives the $0^+$ IAS of $^{48}{\rm Sc}$ at $\varepsilon=7.0$~MeV.
In the reaction calculation the $Q$ value is calculated with the experimental excitation energy 6.7~MeV~\cite{BURROWS20061747} of the IAS.
Note that it is confirmed numerically that the excitation energies of both the TDA and experiment produce identical cross sections.

In addition to the TDA form factor given in Eq.~\eqref{radFF}, we carry out a phenomenological calculation using the Lane model~\cite{Lane1962676},
which is conventionally adopted to compare theoretical charge-exchange cross section values for the Fermi transition with experimental data.
In the Lane model the radial form factor $F_{000}^{s0 {\rm(Lane)}}$ is given as the difference of the optical potentials
between the final and initial channels:
\begin{align}
 % F_{000}^{s0 {\rm(Lane)}}(R)=\frac{A}{2\left(2T_A-1\right)}\left[U_n^{\rm (N)}(R)-U_p^{\rm (N)}(R)\right],
 F_{000}^{s0 {\rm(Lane)}}(R)=\frac{\left(8\pi T_A\right)^{\frac{1}{2}}}{2T_A-1}\left[U_n^{\rm (N)}(R)-U_p^{\rm (N)}(R)\right].
 \label{LaneFF}
\end{align}
where the phenomenological optical potential~\cite{KONING2003231} is used.

\begin{figure}[!t]
\begin{center}
\includegraphics[width=0.48\textwidth,clip]{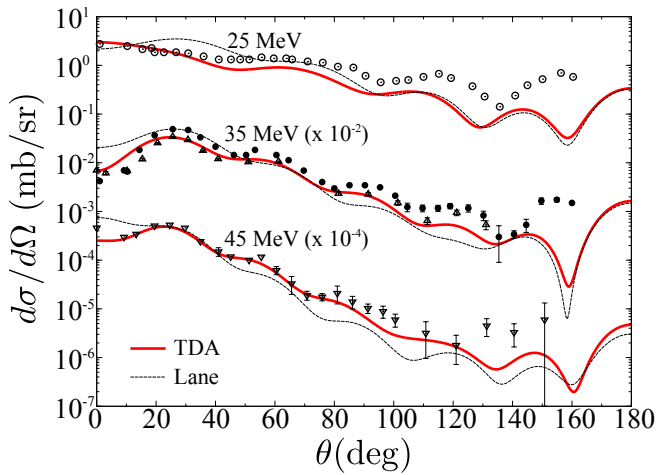}
 \caption{The cross sections of the $^{48}{\rm Ca}(p,n)^{48}{\rm Sc(IAS)}$ reaction at $E_{\rm lab}=$~25,~35,~and~45~MeV.
 The solid lines are the calculated results from the TDA form factor, while the dashed lines are ones from the Lane form factor.
 The measured data are taken from Refs.~\cite{DoeringPhysRevC.12.378,JonPhysRevC.62.044609}.
 The lines and the dots are multiplied by $10^{-2}$ ($10^{-4}$) at 35 (45)~MeV.}
\label{fig1}
\end{center}
\end{figure}

In Fig.~\ref{fig1}, the calculated cross sections of the charge-exchange reaction $^{48}{\rm Ca}(p,n)^{48}{\rm Sc(IAS)}$
at incident proton energies $E_{\rm lab}=$~25, 35, and 45~MeV as a function of the $n$ emitting angle $\theta$ are compared with
experimental data~\cite{DoeringPhysRevC.12.378,JonPhysRevC.62.044609}.
The cross sections calculated by the TDA (Lane) form factor are shown by the solid (dashed) lines.
Note that the theoretical results and experimental data at 35 and 45~MeV are multiplied by $10^{-2}$ and $10^{-4}$, respectively,
in order to make them distinguishable from the cross section at 25~MeV.

%%%%%%%%%%%%%%%
%%% Figure  %%%
%%%%%%%%%%%%%%%

One finds that, in Fig.~\ref{fig1}, the results using the TDA form factor reasonably coincide
with the experimental angular distribution for 35 and 45~MeV.
Although at 25~MeV the TDA result underestimates the data at $\theta \gtrsim 30^\circ$, it appears to be better than the Lane model
in accounting for the measured behavior.
While the Lane model is able to roughly describe the experimental data, it is not as good as the TDA result in the sense of being able to
predict the data.
It should be mentioned that a different choice of optical potential for the Lane model may improve the prediction of the calculation
because its form factor strongly depends on the optical potential used, as reported in Ref.~\cite{KhoaPhysRevC.76.014603}, for example.

%%%%%%%%%%%%%%%%%%%%%%%%%%%%%%%%%%%%%%%%%%%%%%%%%%%%%%%%%%%%%%%%%
\subsection{Spin-flip transitions}
\label{result3}
%%%%%%%%%%%%%%%%%%%%%%%%%%%%%%%%%%%%%%%%%%%%%%%%%%%%%%%%%%%%%%%%%
% \begin{table}[!b]
%  \begin{tabular}{cccc}
%   & $\varepsilon$ (MeV) & $B({\rm GT})$ & $B({\rm SQ})$ \\
% \hline
% \hline
%   \multirow{2}{*}{STDA} & 2.5  & 4.73 & 49.7 \\
%                         & 10.2 & 10.5 & 5.77 \\
% \hline
%   \multirow{2}{*}{TDA} & 3.5  & 5.64 & 49.8 \\
%                        & 11.5 & 18.1 & 14.7 \\
%  \end{tabular}
%  \caption{The $1^+$-resonance energy $\varepsilon$, $B({\rm GT})$, and $B({\rm SQ})$ calculated by STDA (upper rows) and TDA (lower rows).}
%  \label{GTTable}
% \end{table}

We have shown that our framework adequately describes the differential cross section of the $^{48}{\rm Ca}(p,n)^{48}{\rm Sc}({\rm IAS})$ reaction.
Now we switch gears and investigate the 2p2h effect of the $^{48}{\rm Ca}(p,n)^{48}{\rm Sc}(1^+)$ reaction.

% Figure~\ref{strength} shows $B({\rm GT})$ of the low-lying and giant GT resonances as a function of $\varepsilon$.
As seen in Fig.~\ref{strength}(a), the GT strengths manifest themselves in two distant regions: one is around 3~MeV,
which we refer to as the low-lying resonance, and the other is around 11~MeV, which is nothing but the giant GT resonance.
In the STDA, the GT resonance distributes widely due to the 2p2h effect as discussed in Ref.~\cite{MinatoPhysRevC.93.044319}.
Note that we choose the most prominent strength from each region of the low-lying and giant GT resonances
when calculating the differential cross sections.
The strengths of the SQ-$1^+$ transition $B({\rm SQ})$, which are the leading part of the $\Delta l=2$ transition,
are shown in Fig.~\ref{strength}(b).
When we compare cross sections calculated with the STDA and TDA transition densities, the experimental
resonance energy of $\varepsilon=2.6$~MeV (11.0~MeV)~\cite{YakoPhysRevLett.103.012503} is used for the low-lying (giant) resonance.
As in the case of the Fermi transition, this slight shift of the $Q$ value from the theoretical
one does not vary the calculated cross section significantly{; the effect on the cross section at $\theta=0^\circ$ is less than 1\%.

Let us first focus on the low-lying resonance.
Figure~\ref{fig2} shows the differential cross section of the $^{48}{\rm Ca}(p,n)^{48}{\rm Sc}$ reaction at $E_{\rm lab}=$~295~MeV for
the low-lying $1^+$ resonance as a function of $\theta$ up to $40^\circ$.
The cross section calculated by the DWBA with the STDA-transition density is indicated by the solid line,
whereas the cross section calculated with the TDA is represented by the dashed line.
Here the theoretical cross section includes both the GT-type and $\Delta l=2$ transitions.

%%%%%%%%%%%%%%%
%%% Figure  %%%
%%%%%%%%%%%%%%%
\begin{figure}[!t]
\begin{center}
\includegraphics[width=0.48\textwidth,clip]{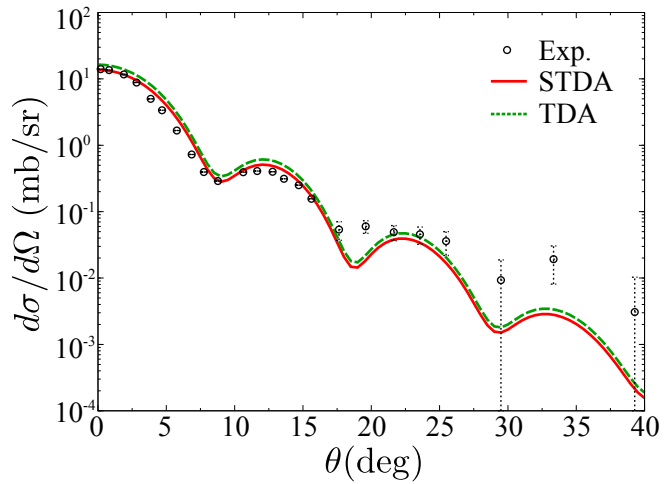}
 \caption{The differential cross section of the $^{48}{\rm Ca}(p,n)^{48}{\rm Sc(GT)}$ reaction
 at $E_{\rm lab}=$~295~MeV for the low-lying $1^+$ resonance state.
 The calculated result with (without) the 2p2h configuration shown by the solid (dashed) line is compared with
 the experimental data (open circle) taken from Ref.~\cite{YakoPhysRevLett.103.012503}.}
\label{fig2}
\end{center}
\end{figure}

Our calculation reproduces the diffraction pattern of the measured cross section reasonably well for both the STDA and TDA.
A difference can be observed only in terms of the magnitude between them.
Using the same value of $\bar{V}_1$ for the TDA and STDA, the cross sections at $\theta=0^\circ$
of the TDA are higher than those of the STDA by about 20\%, and the difference remains almost the same for other angles.

The reductions of the cross section by the 2p2h configuration within the STDA are associated with the reduction of $B({\rm GT})$, $B({\rm SQ})$, and so on.
We obtained $B({\rm GT})=4.726$ for the STDA and 5.681 for the TDA as shown in Fig.~\ref{strength}(a).
The missing strength is brought to a higher energy region \cite{MinatoPhysRevC.93.044319}.
The difference of $B({\rm GT})$ between the TDA and STDA is approximately 20\% and
is equivalent to the reduction due to the 2p2h effect on the cross section.
This proportionality is consistent with the conclusion by Taddeucci {\it et al}.~\cite{Taddeucci1987} although they neglect the $\Delta l=2$ transition.
This fact implies that these contributions are negligibly small (this point will be addressed later).

%%%%%%%%%%%%%%%
%%% Figure  %%%
%%%%%%%%%%%%%%%
\begin{figure}[!t]
\begin{center}
\includegraphics[width=0.48\textwidth,clip]{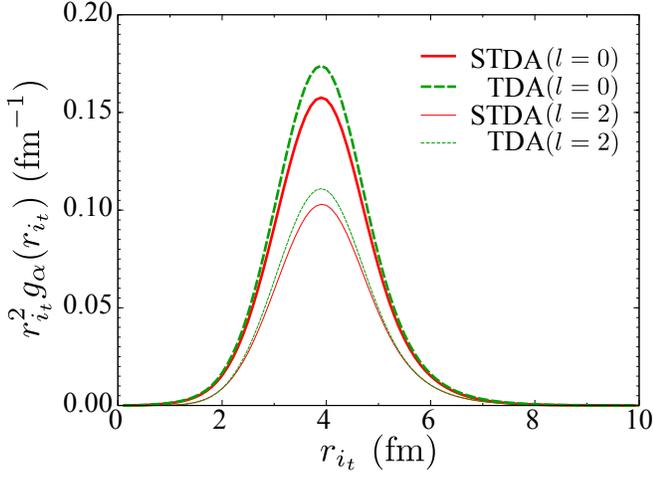}
 \caption{The transition density of the low-lying $1^+$ resonance state of $^{48}{\rm Sc}$
 calculated with the STDA (thick solid line for $l=0$, thin solid line for $l=2$) and TDA (thick dashed line for $l=0$, thin dashed line for $l=2$).}
\label{fig3}
\end{center}
\end{figure}

We plot the transition density $g_{\alpha}$ in Fig.~\ref{fig3} to investigate the difference in the calculated cross sections
of the TDA and STDA in detail.
The thick (thin) solid and thick (thin) dashed lines are respectively the results of the STDA and TDA for $l=0$ ($l=2$)
corresponding to the GT ($\Delta l=2$) transition.
One finds the difference in the amplitudes of the transition density between the STDA and TDA.
Taking the ratio of the STDA amplitude for $l=0$ at the peak around $r_{i_t} \sim 4$~fm with the similar amplitude calculated from the TDA,
we obtain $0.156/0.173\sim0.902$.
Because $B({\rm GT})$ is proportional to $g^2_\alpha$, one obtains $(0.902)^2=0.814$, which is consistent with the reduction of $B({\rm GT})$.

\begin{table}[!b]
 \caption{Leading configurations of the $1^+$ resonance and its amplitude defined by $X_{mi}^2$ of the 1p1h states 
 calculated by the TDA and STDA are listed. The 2p2h amplitude $P_{\rm 2p2h}$ is calculated by
 $P_{\rm 2p2h}=\sum_{mnij}\mathcal{X}_{mnij}^2$.}
 \begin{tabular}{cccc}
                                & Configuration & TDA & STDA \\
  \hline
  \hline
  \multirow{3}{*}{Low-lying GT} &$\pi(1f_{7/2})\nu(1f_{7/2})^{-1}$& 0.954 & 0.858 \\
                                          &$\pi(1f_{5/2})\nu(1f_{7/2})^{-1}$& 0.043 & 0.047 \\
                                          &$\pi(2f_{7/2})\nu(1f_{7/2})^{-1}$& 0.001 & 0.001 \\
                                          &$P_{\rm 2p2h}$              & 0.000 & 0.091 \\
  \hline
  \multirow{3}{*}{Giant GT} &$\pi(1f_{7/2})\nu(1f_{7/2})^{-1}$& 0.042 & 0.043 \\
                                 &$\pi(1f_{5/2})\nu(1f_{7/2})^{-1}$& 0.950 & 0.483 \\
                                 &$\pi(2f_{5/2})\nu(1f_{7/2})^{-1}$& 0.004 & 0.002 \\
                                 &$P_{\rm 2p2h}$              & 0.000 & 0.470 \\
 \end{tabular}
 \label{collectivity}
\end{table}

The diffraction pattern of the cross section has a sensitivity to the shape of the transition density rather than its amplitude
because the angular distribution is determined by the region where the incident proton interacts with the target nucleus.
In Fig.~\ref{fig3}, the STDA and TDA lines have a similar $r_{i_t}$ dependence for each $l$.
Inclusion of the 2p2h configuration does not significantly change the shape of the transition density
although the amplitudes are about 10\% (7\%) smaller for the TDA for $l=0$ $(l=2)$.
In Table~\ref{collectivity}, the 1p1h configurations contributing to the low-lying GT resonance and its amplitude defined by $X_{mi}^2$ are listed.
The main configurations are $\pi(1f_{7/2})\nu(1f_{7/2})^{-1}$ and $\pi(1f_{5/2})\nu(1f_{7/2})^{-1}$
for both the TDA and STDA. While the amplitude of $\pi(1f_{5/2})\nu(1f_{7/2})^{-1}$ is almost the same for both,
the amplitude of $\pi(1f_{7/2})\nu(1f_{7/2})^{-1}$ for the STDA is about 0.1 smaller than that for the TDA.
This difference might change the shape of the transition density if the radial dependences of the wave functions
of $\pi(1f_{7/2})$ and $\pi(1f_{5/2})$ are different.
However, they are almost the same because they are spin-orbit partners.
Therefore, unless another configuration intervenes, the shape of the transition density will not change significantly.
As a consequence, we obtained differential cross sections of similar shape for the STDA and TDA.

Figure~\ref{fig4} shows the cross sections calculated with $g_{s1}$ and $g_{l2}$ (solid line),
only with $g_{s1}$ (dashed line), and only with $g_{l2}$ (dotted line) by means of the STDA,
as well as experimental data~\cite{YakoPhysRevLett.103.012503} (open circle).
Throughout the observed region of $\theta$, the result including only the $\Delta l=2$ transition is about two orders smaller than the others.
At $\theta=0^\circ$, in particular, it is about five orders smaller than that of GT alone
even though $r_{i_t}^2g_{l2}$ has a peak amplitude about 36\% smaller than that of $r_{i_t}^2g_{s1}$ (see Fig.~\ref{fig3}).
It indicates that there are dynamical processes such as angular-momentum coupling coefficients and coherent summation
in Eq.~\eqref{redTmatGen}, which hinder the $\Delta l=2$ components,
and thus the effect of the $\Delta \l=2$ transition on the transition density does not coincide quantitatively with that observed on the cross section.
%%%%%%%%%%%%%%%
%%% Figure  %%%
%%%%%%%%%%%%%%%
\begin{figure}[!t]
\begin{center}
\includegraphics[width=0.48\textwidth,clip]{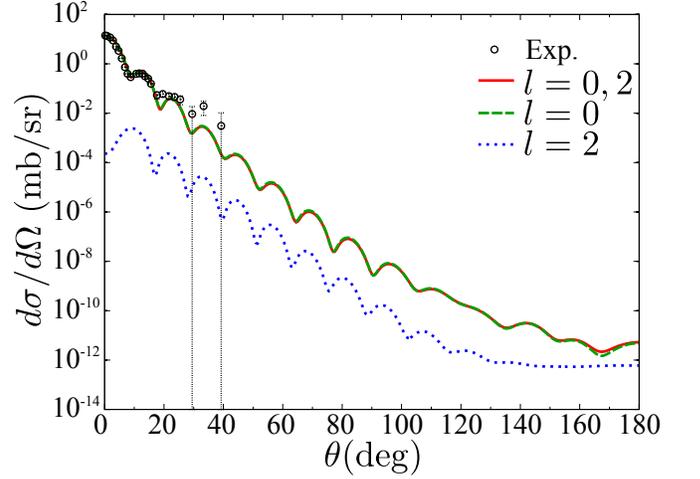}
 \caption{The cross section of $^{48}{\rm Ca}(p,n)^{48}{\rm Sc}$ at 295~MeV for the low-lying $1^+$ resonance state
 calculated with the STDA transition density of the GT and $\Delta l=2$ transitions (solid line), the GT transition only (dashed line),
 and $\Delta l=2$ transition only (dotted line).}
\label{fig4}
\end{center}
\end{figure}

%%%%%%%%%%%%%%%
%%% Figure  %%%
%%%%%%%%%%%%%%%
\begin{figure}[!t]
\begin{center}
\includegraphics[width=0.48\textwidth,clip]{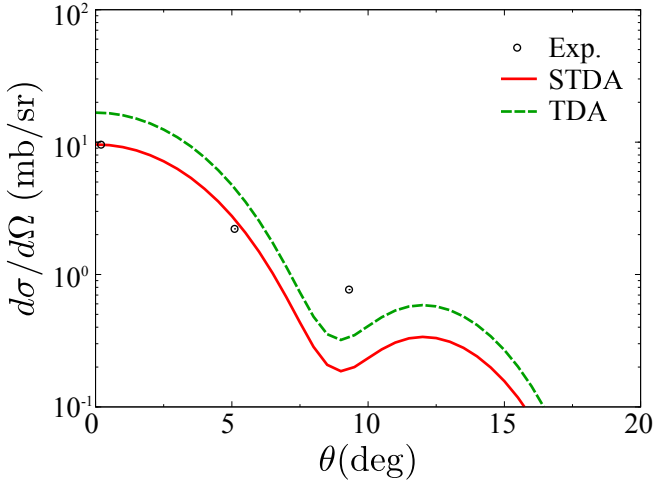}
 \caption{Same as Fig.~\ref{fig2} but for the giant resonance.}
\label{fig5}
\end{center}
\end{figure}

Next we discuss the 2p2h effect on the giant GT resonance.
In Fig.~\ref{fig5} the lines and open circles are defined in the same way as in Fig.~\ref{fig2} but for the giant resonance with $\theta$ up to $20^\circ$.
The result of the STDA reasonably traces the first two points of the experimental data, but fails for the third one.
By the 2p2h effect, the cross section of the STDA is smaller than that of the TDA by about 43\% at $\theta=0^\circ$
but does not change its shape significantly.
Again, comparing $B({\rm GT})$ of the STDA and TDA shown in Fig.~\ref{strength}(a), the 2p2h effect on $B({\rm GT})$ of the giant resonance
is about a 42\% reduction, which agrees with the value of its effect on the cross section.
%%%%%%%%%%%%%%%
%%% Figure  %%%
%%%%%%%%%%%%%%%
\begin{figure}[!b]
\begin{center}
\includegraphics[width=0.48\textwidth,clip]{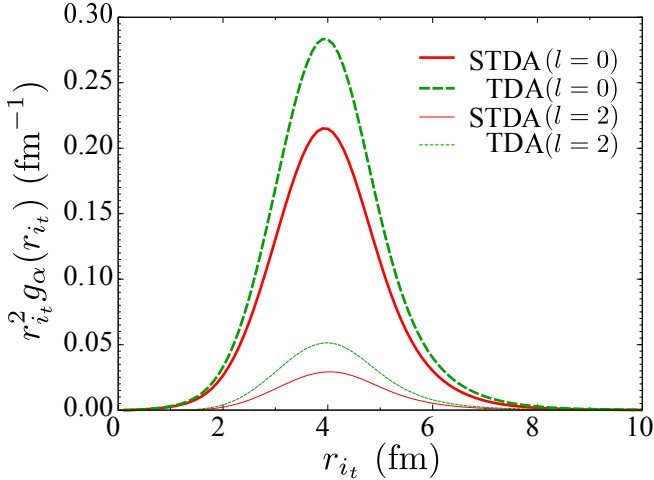}
 \caption{Same as Fig.~\ref{fig3} but for the giant resonance.}
\label{fig6}
\end{center}
\end{figure}

Figure~\ref{fig6} shows the transition density of the giant resonance. From the difference between the STDA and TDA,
we find that the 2p2h configuration reduces the amplitude of $r_{i_t}^2g_{\alpha}$ at the peak position around $r_{i_t}=4$~fm
by about 25\% (43\%) for $l=0$ ($l=2$). As we did in the low-lying resonance, calculating the squared ratio of the amplitude of
the STDA to that of the TDA,
one obtains $(0.211/0.281)^2\sim0.564$, which is almost consistent with the reductions of $B(\rm{GT})$ and the cross section.
From Table~\ref{collectivity}, the 1p1h configurations mainly contributing to the giant GT resonance are $\pi(1f_{7/2})\nu(1f_{7/2})^{-1}$
and $\pi(1f_{5/2})\nu(1f_{7/2})^{-1}$ both for the TDA and STDA, as in the case of the low-lying resonance.
While the amplitude of $\pi(1f_{7/2})\nu(1f_{7/2})^{-1}$ almost remains the same for both the TDA and STDA, that of $\pi(1f_{5/2})\nu(1f_{7/2})^{-1}$
for the STDA is half of that for the TDA.
However, this difference does not make a significant change in the shape of the transition density and accordingly
in the diffraction pattern of the cross section, similar to the low-lying resonance, as seen in Fig.~\ref{fig5}.

%%%%%%%%%%%%%%%
%%% Figure  %%%
%%%%%%%%%%%%%%%
\begin{figure}[!t]
\begin{center}
\includegraphics[width=0.48\textwidth,clip]{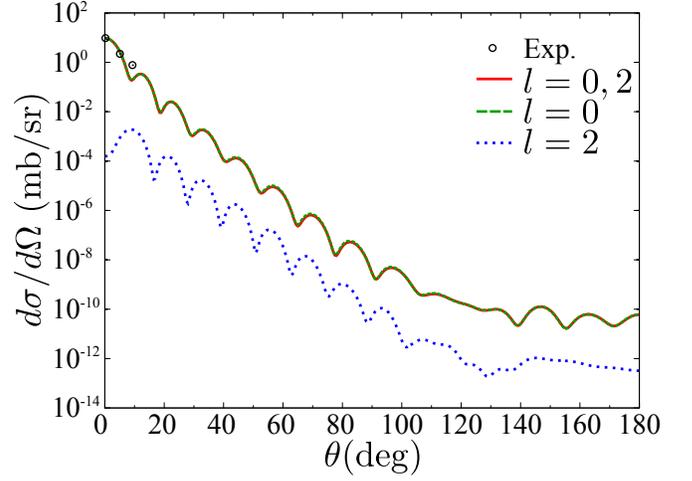}
 \caption{Same as Fig.~\ref{fig4} but for the giant resonance.}
\label{fig7}
\end{center}
\end{figure}
Figure~\ref{fig7} shows the cross section at the giant GT resonance.
The result of $\Delta l=2$ transition only is negligibly small as compared to the others.
It is about two orders smaller than the GT transition in $\theta > 0$, and  the ratio of their cross sections
at $\theta=0^\circ$ is approximately $10^{-5} $, similar to the result of the low-lying resonance.

As a consequence, qualitatively the 2p2h effect reduces the amplitude of the cross section but does not change the diffraction pattern.
% Quantitatively the 2p2h effect is more significant for the giant resonance than low-energy one.
The values of the decrease on the cross section due to the 2p2h configuration are
essentially consistent with those obtained from the structural calculation.

Last, we comment on the tensor-force contribution, which was reported~\cite{MinatoPhysRevC.93.044319} to change
the excitation energy of the spin-flip resonance states and the corresponding $B({\rm GT})$ values.
However, we have confirmed numerically that the inclusion of the tensor force does not change the diffraction pattern of the cross section.
%conclusion that the 2p2h effect on the cross section is directly linked to that on the transition density.

%%%%%%%%%%%%%%%%%%%%%%%%%%%%%%%%%%%%%%%%%%%%%%%%%%%%%%%%%%%%%%%%%
\section{Summary}
\label{summary}
%%%%%%%%%%%%%%%%%%%%%%%%%%%%%%%%%%%%%%%%%%%%%%%%%%%%%%%%%%%%%%%%%
The charge-exchange reaction $^{48}{\rm Ca}(p,n)^{48}{\rm Sc}$ has been investigated theoretically to clarify
the effect of 2p2h-configuration mixing on the GT-resonance states.
We have carried out the STDA calculation in order to prepare the transition density,
and the form factor has been obtained by employing the phenomenological nucleon-nucleon interaction.
The angular-distributed cross section has been computed by means of the DWBA with the microscopic form factor.

The Fermi transition has also been calculated to demonstrate the effectiveness of our framework.
The calculated cross sections of the Fermi transition caused by the $^{48}{\rm Ca}(p,n)^{48}{\rm Sc(IAS)}$ reaction at $E_{\rm lab}=$~25, 35, and 45~MeV
coincide well with the measured data~\cite{DoeringPhysRevC.12.378,JonPhysRevC.62.044609}.

It has been found that the 2p2h effect on the cross section of the $^{48}{\rm Ca}(p,n)^{48}{\rm Sc}$ reaction at 295~MeV
decreases the amplitude of the cross section and does not change the angular distribution for either the low-lying or giant resonances.
This feature is consistent with the result of the structural calculation.
However, the 2p2h effect on the angular distribution may become important for other multipole transitions
because it was reported that the transition densities of the isovector monopole and the quadrupole of $^{16}$O
were changed significantly~\cite{Phys.Rev.C81.054312}.
Quantitatively, the reduction of the cross section due to the 2p2h effect can be explained by that of $B({\rm GT})$
and the corresponding transition density.

The role of the $\Delta l=2$ transition on $1^+$-resonance states has also been surveyed and
found to give a negligibly small contribution.
It supports the proportion relation~\cite{Taddeucci1987} between $B({\rm GT})$ and the charge-exchange cross section at zero degree.
Note that, in our model, the form factor of the $\Delta l=2$ transition has been calculated using the same nucleon-nucleon interaction
as that of the GT transition.
A different nucleon-nucleon interaction should be tested, for example,
the $t$ matrix of Franey and Love~\cite{FLPhysRevC.31.488}
or the $g$ matrix of Jeukenne-Lejeune-Mahaux~\cite{JeukennePhysRevC.16.80},
as adopted in previous studies~\cite{Taddeucci1987,KERMAN1959551,BE0034-4885-50-6-001,KhoaPhysRevC.76.014603}.

A systematic comparison of the reaction models such as the DWBA, DWIA, and coupled-channels method
for the charge-exchange reaction at several incident energies with several target nuclei
will provide important guidance for analyses of experimental data.

%----------------------------------------------
 \begin{acknowledgments}
  The authors thank K.~Hagino, O.~Iwamoto, and K.~Minomo for constructive comments and suggestions.
  They also thank E.~Olsen for helpful advice and refining our discussion.
 \end{acknowledgments}
%----------------------------------------------

\nocite{*}

%+++++++++++++++++++++++++++++++++++++++++++++++++++++++++++++++++++++
\bibliography{ChargeEX}

\begin{thebibliography}{40}%
\makeatletter
\providecommand \@ifxundefined [1]{%
 \@ifx{#1\undefined}
}%
\providecommand \@ifnum [1]{%
 \ifnum #1\expandafter \@firstoftwo
 \else \expandafter \@secondoftwo
 \fi
}%
\providecommand \@ifx [1]{%
 \ifx #1\expandafter \@firstoftwo
 \else \expandafter \@secondoftwo
 \fi
}%
\providecommand \natexlab [1]{#1}%
\providecommand \enquote  [1]{``#1''}%
\providecommand \bibnamefont  [1]{#1}%
\providecommand \bibfnamefont [1]{#1}%
\providecommand \citenamefont [1]{#1}%
\providecommand \href@noop [0]{\@secondoftwo}%
\providecommand \href [0]{\begingroup \@sanitize@url \@href}%
\providecommand \@href[1]{\@@startlink{#1}\@@href}%
\providecommand \@@href[1]{\endgroup#1\@@endlink}%
\providecommand \@sanitize@url [0]{\catcode `\\12\catcode `\$12\catcode
  `\&12\catcode `\#12\catcode `\^12\catcode `\_12\catcode `\%12\relax}%
\providecommand \@@startlink[1]{}%
\providecommand \@@endlink[0]{}%
\providecommand \url  [0]{\begingroup\@sanitize@url \@url }%
\providecommand \@url [1]{\endgroup\@href {#1}{\urlprefix }}%
\providecommand \urlprefix  [0]{URL }%
\providecommand \Eprint [0]{\href }%
\providecommand \doibase [0]{http://dx.doi.org/}%
\providecommand \selectlanguage [0]{\@gobble}%
\providecommand \bibinfo  [0]{\@secondoftwo}%
\providecommand \bibfield  [0]{\@secondoftwo}%
\providecommand \translation [1]{[#1]}%
\providecommand \BibitemOpen [0]{}%
\providecommand \bibitemStop [0]{}%
\providecommand \bibitemNoStop [0]{.\EOS\space}%
\providecommand \EOS [0]{\spacefactor3000\relax}%
\providecommand \BibitemShut  [1]{\csname bibitem#1\endcsname}%
\let\auto@bib@innerbib\@empty
%</preamble>
\bibitem [{\citenamefont {Martini}\ \emph {et~al.}(2009)\citenamefont
  {Martini}, \citenamefont {Ericson}, \citenamefont {Chanfray},\ and\
  \citenamefont {Marteau}}]{MartiniPhysRevC.80.065501}%
  \BibitemOpen
  \bibfield  {author} {\bibinfo {author} {\bibfnamefont {M.}~\bibnamefont
  {Martini}}, \bibinfo {author} {\bibfnamefont {M.}~\bibnamefont {Ericson}},
  \bibinfo {author} {\bibfnamefont {G.}~\bibnamefont {Chanfray}}, \ and\
  \bibinfo {author} {\bibfnamefont {J.}~\bibnamefont {Marteau}},\ }\href
  {\doibase 10.1103/PhysRevC.80.065501} {\bibfield  {journal} {\bibinfo
  {journal} {Phys. Rev. C}\ }\textbf {\bibinfo {volume} {80}},\ \bibinfo
  {pages} {065501} (\bibinfo {year} {2009})}\BibitemShut {NoStop}%
\bibitem [{\citenamefont {Ikeda}(1964)}]{Ikeda01031964}%
  \BibitemOpen
  \bibfield  {author} {\bibinfo {author} {\bibfnamefont {K.}~\bibnamefont
  {Ikeda}},\ }\href {\doibase 10.1143/PTP.31.434} {\bibfield  {journal}
  {\bibinfo  {journal} {Prog. Theor, Phys.}\ }\textbf {\bibinfo {volume}
  {31}},\ \bibinfo {pages} {434} (\bibinfo {year} {1964})}\BibitemShut
  {NoStop}%
\bibitem [{\citenamefont {Fujita}\ and\ \citenamefont
  {Ikeda}(1965)}]{FUJITA1965145}%
  \BibitemOpen
  \bibfield  {author} {\bibinfo {author} {\bibfnamefont {J.-I.}\ \bibnamefont
  {Fujita}}\ and\ \bibinfo {author} {\bibfnamefont {K.}~\bibnamefont {Ikeda}},\
  }\href {\doibase http://dx.doi.org/10.1016/0029-5582(65)90119-7} {\bibfield
  {journal} {\bibinfo  {journal} {Nucl. Phys.}\ }\textbf {\bibinfo {volume}
  {67}},\ \bibinfo {pages} {145 } (\bibinfo {year} {1965})}\BibitemShut
  {NoStop}%
\bibitem [{\citenamefont {Gaarde}(1983)}]{Gaarde1983}%
  \BibitemOpen
  \bibfield  {author} {\bibinfo {author} {\bibfnamefont {C.}~\bibnamefont
  {Gaarde}},\ }\href@noop {} {\bibfield  {journal} {\bibinfo  {journal} {Nucl.
  Phys.}\ }\textbf {\bibinfo {volume} {A396}},\ \bibinfo {pages} {127}
  (\bibinfo {year} {1983})}\BibitemShut {NoStop}%
\bibitem [{\citenamefont {Bai}\ \emph {et~al.}(2009)\citenamefont {Bai},
  \citenamefont {Zhang}, \citenamefont {Zhang}, \citenamefont {Xu},
  \citenamefont {Sagawa},\ and\ \citenamefont {Col\`{o}}}]{BAI2009}%
  \BibitemOpen
  \bibfield  {author} {\bibinfo {author} {\bibfnamefont {C.}~\bibnamefont
  {Bai}}, \bibinfo {author} {\bibfnamefont {H.}~\bibnamefont {Zhang}}, \bibinfo
  {author} {\bibfnamefont {X.}~\bibnamefont {Zhang}}, \bibinfo {author}
  {\bibfnamefont {F.}~\bibnamefont {Xu}}, \bibinfo {author} {\bibfnamefont
  {H.}~\bibnamefont {Sagawa}}, \ and\ \bibinfo {author} {\bibfnamefont
  {G.}~\bibnamefont {Col\`{o}}},\ }\href {\doibase
  http://dx.doi.org/10.1103/PhysRevC.79.041301} {\bibfield  {journal} {\bibinfo
   {journal} {Phys. Rev. C}\ }\textbf {\bibinfo {volume} {79}},\ \bibinfo
  {pages} {041301(R)} (\bibinfo {year} {2009})}\BibitemShut {NoStop}%
\bibitem [{\citenamefont {Ichimura}\ \emph {et~al.}(2006)\citenamefont
  {Ichimura}, \citenamefont {Sakai},\ and\ \citenamefont
  {Wakasa}}]{Prog.Part.Nucl.Phys.56.446}%
  \BibitemOpen
  \bibfield  {author} {\bibinfo {author} {\bibfnamefont {M.}~\bibnamefont
  {Ichimura}}, \bibinfo {author} {\bibfnamefont {H.}~\bibnamefont {Sakai}}, \
  and\ \bibinfo {author} {\bibfnamefont {T.}~\bibnamefont {Wakasa}},\
  }\href@noop {} {\bibfield  {journal} {\bibinfo  {journal} {Prog. Part. Nucl.
  Phys.}\ }\textbf {\bibinfo {volume} {56}},\ \bibinfo {pages} {446} (\bibinfo
  {year} {2006})}\BibitemShut {NoStop}%
\bibitem [{\citenamefont {Minato}(2016)}]{MinatoPhysRevC.93.044319}%
  \BibitemOpen
  \bibfield  {author} {\bibinfo {author} {\bibfnamefont {F.}~\bibnamefont
  {Minato}},\ }\href {\doibase 10.1103/PhysRevC.93.044319} {\bibfield
  {journal} {\bibinfo  {journal} {Phys. Rev. C}\ }\textbf {\bibinfo {volume}
  {93}},\ \bibinfo {pages} {044319} (\bibinfo {year} {2016})}\BibitemShut
  {NoStop}%
\bibitem [{\citenamefont {Yako}\ \emph {et~al.}(2009)\citenamefont {Yako},
  \citenamefont {Sasano}, \citenamefont {Miki}, \citenamefont {Sakai},
  \citenamefont {Dozono}, \citenamefont {Frekers}, \citenamefont {Greenfield},
  \citenamefont {Hatanaka}, \citenamefont {Ihara}, \citenamefont {Kato},
  \citenamefont {Kawabata}, \citenamefont {Kuboki}, \citenamefont {Maeda},
  \citenamefont {Matsubara}, \citenamefont {Muto}, \citenamefont {Noji},
  \citenamefont {Okamura}, \citenamefont {Okabe}, \citenamefont {Sakaguchi},
  \citenamefont {Sakemi}, \citenamefont {Sasamoto}, \citenamefont {Sekiguchi},
  \citenamefont {Shimizu}, \citenamefont {Suda}, \citenamefont {Tameshige},
  \citenamefont {Tamii}, \citenamefont {Uesaka}, \citenamefont {Wakasa},\ and\
  \citenamefont {Zheng}}]{YakoPhysRevLett.103.012503}%
  \BibitemOpen
  \bibfield  {author} {\bibinfo {author} {\bibfnamefont {K.}~\bibnamefont
  {Yako}}, \bibinfo {author} {\bibfnamefont {M.}~\bibnamefont {Sasano}},
  \bibinfo {author} {\bibfnamefont {K.}~\bibnamefont {Miki}}, \bibinfo {author}
  {\bibfnamefont {H.}~\bibnamefont {Sakai}}, \bibinfo {author} {\bibfnamefont
  {M.}~\bibnamefont {Dozono}}, \bibinfo {author} {\bibfnamefont
  {D.}~\bibnamefont {Frekers}}, \bibinfo {author} {\bibfnamefont {M.~B.}\
  \bibnamefont {Greenfield}}, \bibinfo {author} {\bibfnamefont
  {K.}~\bibnamefont {Hatanaka}}, \bibinfo {author} {\bibfnamefont
  {E.}~\bibnamefont {Ihara}}, \bibinfo {author} {\bibfnamefont
  {M.}~\bibnamefont {Kato}}, \bibinfo {author} {\bibfnamefont {T.}~\bibnamefont
  {Kawabata}}, \bibinfo {author} {\bibfnamefont {H.}~\bibnamefont {Kuboki}},
  \bibinfo {author} {\bibfnamefont {Y.}~\bibnamefont {Maeda}}, \bibinfo
  {author} {\bibfnamefont {H.}~\bibnamefont {Matsubara}}, \bibinfo {author}
  {\bibfnamefont {K.}~\bibnamefont {Muto}}, \bibinfo {author} {\bibfnamefont
  {S.}~\bibnamefont {Noji}}, \bibinfo {author} {\bibfnamefont {H.}~\bibnamefont
  {Okamura}}, \bibinfo {author} {\bibfnamefont {T.~H.}\ \bibnamefont {Okabe}},
  \bibinfo {author} {\bibfnamefont {S.}~\bibnamefont {Sakaguchi}}, \bibinfo
  {author} {\bibfnamefont {Y.}~\bibnamefont {Sakemi}}, \bibinfo {author}
  {\bibfnamefont {Y.}~\bibnamefont {Sasamoto}}, \bibinfo {author}
  {\bibfnamefont {K.}~\bibnamefont {Sekiguchi}}, \bibinfo {author}
  {\bibfnamefont {Y.}~\bibnamefont {Shimizu}}, \bibinfo {author} {\bibfnamefont
  {K.}~\bibnamefont {Suda}}, \bibinfo {author} {\bibfnamefont {Y.}~\bibnamefont
  {Tameshige}}, \bibinfo {author} {\bibfnamefont {A.}~\bibnamefont {Tamii}},
  \bibinfo {author} {\bibfnamefont {T.}~\bibnamefont {Uesaka}}, \bibinfo
  {author} {\bibfnamefont {T.}~\bibnamefont {Wakasa}}, \ and\ \bibinfo {author}
  {\bibfnamefont {H.}~\bibnamefont {Zheng}},\ }\href {\doibase
  10.1103/PhysRevLett.103.012503} {\bibfield  {journal} {\bibinfo  {journal}
  {Phys. Rev. Lett.}\ }\textbf {\bibinfo {volume} {103}},\ \bibinfo {pages}
  {012503} (\bibinfo {year} {2009})}\BibitemShut {NoStop}%
\bibitem [{\citenamefont {Dro\.{z}d\.{z}}\ \emph {et~al.}(1986)\citenamefont
  {Dro\.{z}d\.{z}}, \citenamefont {Klemt}, \citenamefont {Speth},\ and\
  \citenamefont {Wambach}}]{Phys.Lett.166B.18}%
  \BibitemOpen
  \bibfield  {author} {\bibinfo {author} {\bibfnamefont {S.}~\bibnamefont
  {Dro\.{z}d\.{z}}}, \bibinfo {author} {\bibfnamefont {V.}~\bibnamefont
  {Klemt}}, \bibinfo {author} {\bibfnamefont {J.}~\bibnamefont {Speth}}, \ and\
  \bibinfo {author} {\bibfnamefont {J.}~\bibnamefont {Wambach}},\ }\href@noop
  {} {\bibfield  {journal} {\bibinfo  {journal} {Phys. Lett. B}\ }\textbf
  {\bibinfo {volume} {166}},\ \bibinfo {pages} {18} (\bibinfo {year}
  {1986})}\BibitemShut {NoStop}%
\bibitem [{\citenamefont {Niu}\ \emph {et~al.}(2014)\citenamefont {Niu},
  \citenamefont {Col\`{o}},\ and\ \citenamefont
  {Vigezzi}}]{Phys.Rev.C90.054328}%
  \BibitemOpen
  \bibfield  {author} {\bibinfo {author} {\bibfnamefont {Y.~F.}\ \bibnamefont
  {Niu}}, \bibinfo {author} {\bibfnamefont {G.}~\bibnamefont {Col\`{o}}}, \
  and\ \bibinfo {author} {\bibfnamefont {E.}~\bibnamefont {Vigezzi}},\
  }\href@noop {} {\bibfield  {journal} {\bibinfo  {journal} {Phys. Rev. C}\
  }\textbf {\bibinfo {volume} {90}},\ \bibinfo {pages} {054328} (\bibinfo
  {year} {2014})}\BibitemShut {NoStop}%
\bibitem [{\citenamefont {Wakasa}\ \emph {et~al.}(2012)\citenamefont {Wakasa},
  \citenamefont {Okamoto}, \citenamefont {Dozono}, \citenamefont {Hatanaka},
  \citenamefont {Ichimura}, \citenamefont {Kuroita}, \citenamefont {Maeda},
  \citenamefont {Miyasako}, \citenamefont {Noro}, \citenamefont {Saito},
  \citenamefont {Sakemi}, \citenamefont {Yabe},\ and\ \citenamefont
  {Yako}}]{Wakasa2012}%
  \BibitemOpen
  \bibfield  {author} {\bibinfo {author} {\bibfnamefont {T.}~\bibnamefont
  {Wakasa}}, \bibinfo {author} {\bibfnamefont {M.}~\bibnamefont {Okamoto}},
  \bibinfo {author} {\bibfnamefont {M.}~\bibnamefont {Dozono}}, \bibinfo
  {author} {\bibfnamefont {K.}~\bibnamefont {Hatanaka}}, \bibinfo {author}
  {\bibfnamefont {M.}~\bibnamefont {Ichimura}}, \bibinfo {author}
  {\bibfnamefont {S.}~\bibnamefont {Kuroita}}, \bibinfo {author} {\bibfnamefont
  {Y.}~\bibnamefont {Maeda}}, \bibinfo {author} {\bibfnamefont
  {H.}~\bibnamefont {Miyasako}}, \bibinfo {author} {\bibfnamefont
  {T.}~\bibnamefont {Noro}}, \bibinfo {author} {\bibfnamefont {T.}~\bibnamefont
  {Saito}}, \bibinfo {author} {\bibfnamefont {Y.}~\bibnamefont {Sakemi}},
  \bibinfo {author} {\bibfnamefont {T.}~\bibnamefont {Yabe}}, \ and\ \bibinfo
  {author} {\bibfnamefont {K.}~\bibnamefont {Yako}},\ }\href {\doibase
  http://dx.doi.org/10.1103/PhysRevC.85.064606} {\bibfield  {journal} {\bibinfo
   {journal} {Phys. Rev. C}\ }\textbf {\bibinfo {volume} {85}},\ \bibinfo
  {pages} {064606} (\bibinfo {year} {2012})}\BibitemShut {NoStop}%
\bibitem [{\citenamefont {Wambach}(1988)}]{Wambach1988}%
  \BibitemOpen
  \bibfield  {author} {\bibinfo {author} {\bibfnamefont {J.}~\bibnamefont
  {Wambach}},\ }\href@noop {} {\bibfield  {journal} {\bibinfo  {journal} {Rep.
  Prog. Phys.}\ }\textbf {\bibinfo {volume} {51}},\ \bibinfo {pages} {989}
  (\bibinfo {year} {1988})}\BibitemShut {NoStop}%
\bibitem [{\citenamefont {Taddeucci}\ \emph {et~al.}(1987)\citenamefont
  {Taddeucci}, \citenamefont {Goulding}, \citenamefont {Carey}, \citenamefont
  {Byrd}, \citenamefont {Goodman}, \citenamefont {Gaarde}, \citenamefont
  {Larsen}, \citenamefont {Horen}, \citenamefont {Raraport},\ and\
  \citenamefont {Sugarbaker}}]{Taddeucci1987}%
  \BibitemOpen
  \bibfield  {author} {\bibinfo {author} {\bibfnamefont {T.}~\bibnamefont
  {Taddeucci}}, \bibinfo {author} {\bibfnamefont {C.}~\bibnamefont {Goulding}},
  \bibinfo {author} {\bibfnamefont {T.}~\bibnamefont {Carey}}, \bibinfo
  {author} {\bibfnamefont {R.}~\bibnamefont {Byrd}}, \bibinfo {author}
  {\bibfnamefont {C.}~\bibnamefont {Goodman}}, \bibinfo {author} {\bibfnamefont
  {C.}~\bibnamefont {Gaarde}}, \bibinfo {author} {\bibfnamefont
  {J.}~\bibnamefont {Larsen}}, \bibinfo {author} {\bibfnamefont
  {D.}~\bibnamefont {Horen}}, \bibinfo {author} {\bibfnamefont
  {J.}~\bibnamefont {Raraport}}, \ and\ \bibinfo {author} {\bibfnamefont
  {E.}~\bibnamefont {Sugarbaker}},\ }\href@noop {} {\bibfield  {journal}
  {\bibinfo  {journal} {Nucl. Phys.}\ }\textbf {\bibinfo {volume} {A469}},\
  \bibinfo {pages} {125} (\bibinfo {year} {1987})}\BibitemShut {NoStop}%
\bibitem [{\citenamefont {Anderson}\ \emph {et~al.}(1985)\citenamefont
  {Anderson}, \citenamefont {Chittrakarn}, \citenamefont {Baldwin},
  \citenamefont {Lebo}, \citenamefont {Madey}, \citenamefont {McCarthy},
  \citenamefont {Watson}, \citenamefont {Brown},\ and\ \citenamefont
  {Foster}}]{PhysRevC.31.1147}%
  \BibitemOpen
  \bibfield  {author} {\bibinfo {author} {\bibfnamefont {B.~D.}\ \bibnamefont
  {Anderson}}, \bibinfo {author} {\bibfnamefont {T.}~\bibnamefont
  {Chittrakarn}}, \bibinfo {author} {\bibfnamefont {A.~R.}\ \bibnamefont
  {Baldwin}}, \bibinfo {author} {\bibfnamefont {C.}~\bibnamefont {Lebo}},
  \bibinfo {author} {\bibfnamefont {R.}~\bibnamefont {Madey}}, \bibinfo
  {author} {\bibfnamefont {R.~J.}\ \bibnamefont {McCarthy}}, \bibinfo {author}
  {\bibfnamefont {J.~W.}\ \bibnamefont {Watson}}, \bibinfo {author}
  {\bibfnamefont {B.~A.}\ \bibnamefont {Brown}}, \ and\ \bibinfo {author}
  {\bibfnamefont {C.~C.}\ \bibnamefont {Foster}},\ }\href {\doibase
  10.1103/PhysRevC.31.1147} {\bibfield  {journal} {\bibinfo  {journal} {Phys.
  Rev. C}\ }\textbf {\bibinfo {volume} {31}},\ \bibinfo {pages} {1147}
  (\bibinfo {year} {1985})}\BibitemShut {NoStop}%
\bibitem [{\citenamefont {Love}\ \emph {et~al.}(1987)\citenamefont {Love},
  \citenamefont {Nakayama},\ and\ \citenamefont
  {Franey}}]{Phys.Rev.Lett.59.1401}%
  \BibitemOpen
  \bibfield  {author} {\bibinfo {author} {\bibfnamefont {W.}~\bibnamefont
  {Love}}, \bibinfo {author} {\bibfnamefont {K.}~\bibnamefont {Nakayama}}, \
  and\ \bibinfo {author} {\bibfnamefont {M.}~\bibnamefont {Franey}},\
  }\href@noop {} {\bibfield  {journal} {\bibinfo  {journal} {Phys. Rev. Lett.}\
  }\textbf {\bibinfo {volume} {59}},\ \bibinfo {pages} {1401} (\bibinfo {year}
  {1987})}\BibitemShut {NoStop}%
\bibitem [{\citenamefont {Fujita}\ \emph {et~al.}(2015)\citenamefont {Fujita},
  \citenamefont {Fujita}, \citenamefont {Adachi}, \citenamefont {Susoy},
  \citenamefont {Algora}, \citenamefont {Bai}, \citenamefont {Col\`{o}},
  \citenamefont {Csatl\'{o}s}, \citenamefont {Deaven}, \citenamefont
  {Estevez-Aguado}, \citenamefont {Guess}, \citenamefont {Guly\'{a}s},
  \citenamefont {Hatanaka}, \citenamefont {Hirota}, \citenamefont {Honma},
  \citenamefont {Ishikawa}, \citenamefont {Krasznahorkay}, \citenamefont
  {Matsubara}, \citenamefont {Meharchand}, \citenamefont {Molina},
  \citenamefont {Nakada}, \citenamefont {Okamura}, \citenamefont {Ong},
  \citenamefont {Otsuka}, \citenamefont {Perdikakis}, \citenamefont {Rubio},
  \citenamefont {Sagawa}, \citenamefont {Sarriguren}, \citenamefont {Scholl},
  \citenamefont {Shimbara}, \citenamefont {Stephenson}, \citenamefont {Tamii},
  \citenamefont {Thies}, \citenamefont {Yoshida}, \citenamefont {Zegers},\ and\
  \citenamefont {Zenihiro}}]{Phys.Rev.C91.064316}%
  \BibitemOpen
  \bibfield  {author} {\bibinfo {author} {\bibfnamefont {Y.}~\bibnamefont
  {Fujita}}, \bibinfo {author} {\bibfnamefont {H.}~\bibnamefont {Fujita}},
  \bibinfo {author} {\bibfnamefont {T.}~\bibnamefont {Adachi}}, \bibinfo
  {author} {\bibfnamefont {G.}~\bibnamefont {Susoy}}, \bibinfo {author}
  {\bibfnamefont {A.}~\bibnamefont {Algora}}, \bibinfo {author} {\bibfnamefont
  {C.}~\bibnamefont {Bai}}, \bibinfo {author} {\bibfnamefont {G.}~\bibnamefont
  {Col\`{o}}}, \bibinfo {author} {\bibfnamefont {M.}~\bibnamefont
  {Csatl\'{o}s}}, \bibinfo {author} {\bibfnamefont {J.}~\bibnamefont {Deaven}},
  \bibinfo {author} {\bibfnamefont {E.}~\bibnamefont {Estevez-Aguado}},
  \bibinfo {author} {\bibfnamefont {C.}~\bibnamefont {Guess}}, \bibinfo
  {author} {\bibfnamefont {J.}~\bibnamefont {Guly\'{a}s}}, \bibinfo {author}
  {\bibfnamefont {K.}~\bibnamefont {Hatanaka}}, \bibinfo {author}
  {\bibfnamefont {K.}~\bibnamefont {Hirota}}, \bibinfo {author} {\bibfnamefont
  {M.}~\bibnamefont {Honma}}, \bibinfo {author} {\bibfnamefont
  {D.}~\bibnamefont {Ishikawa}}, \bibinfo {author} {\bibfnamefont
  {A.}~\bibnamefont {Krasznahorkay}}, \bibinfo {author} {\bibfnamefont
  {H.}~\bibnamefont {Matsubara}}, \bibinfo {author} {\bibfnamefont
  {R.}~\bibnamefont {Meharchand}}, \bibinfo {author} {\bibfnamefont
  {F.}~\bibnamefont {Molina}}, \bibinfo {author} {\bibfnamefont
  {H.}~\bibnamefont {Nakada}}, \bibinfo {author} {\bibfnamefont
  {H.}~\bibnamefont {Okamura}}, \bibinfo {author} {\bibfnamefont
  {H.}~\bibnamefont {Ong}}, \bibinfo {author} {\bibfnamefont {T.}~\bibnamefont
  {Otsuka}}, \bibinfo {author} {\bibfnamefont {G.}~\bibnamefont {Perdikakis}},
  \bibinfo {author} {\bibfnamefont {B.}~\bibnamefont {Rubio}}, \bibinfo
  {author} {\bibfnamefont {H.}~\bibnamefont {Sagawa}}, \bibinfo {author}
  {\bibfnamefont {P.}~\bibnamefont {Sarriguren}}, \bibinfo {author}
  {\bibfnamefont {C.}~\bibnamefont {Scholl}}, \bibinfo {author} {\bibfnamefont
  {Y.}~\bibnamefont {Shimbara}}, \bibinfo {author} {\bibfnamefont {S.~T.}\
  \bibnamefont {Stephenson}, \bibfnamefont {E.J.~and}}, \bibinfo {author}
  {\bibfnamefont {A.}~\bibnamefont {Tamii}}, \bibinfo {author} {\bibfnamefont
  {J.}~\bibnamefont {Thies}}, \bibinfo {author} {\bibfnamefont
  {K.}~\bibnamefont {Yoshida}}, \bibinfo {author} {\bibfnamefont
  {R.}~\bibnamefont {Zegers}}, \ and\ \bibinfo {author} {\bibfnamefont
  {J.}~\bibnamefont {Zenihiro}},\ }\href@noop {} {\bibfield  {journal}
  {\bibinfo  {journal} {Phys. Rev. C}\ }\textbf {\bibinfo {volume} {91}},\
  \bibinfo {pages} {064316} (\bibinfo {year} {2015})}\BibitemShut {NoStop}%
\bibitem [{\citenamefont {Vautherin}\ and\ \citenamefont
  {Brink}(1972)}]{VautherinBrink}%
  \BibitemOpen
  \bibfield  {author} {\bibinfo {author} {\bibfnamefont {D.}~\bibnamefont
  {Vautherin}}\ and\ \bibinfo {author} {\bibfnamefont {D.}~\bibnamefont
  {Brink}},\ }\href@noop {} {\bibfield  {journal} {\bibinfo  {journal} {Phys.
  Rev. C}\ }\textbf {\bibinfo {volume} {5}},\ \bibinfo {pages} {626} (\bibinfo
  {year} {1972})}\BibitemShut {NoStop}%
\bibitem [{\citenamefont {Nishizaki}\ \emph {et~al.}(1988)\citenamefont
  {Nishizaki}, \citenamefont {Dro\.{z}d\.{z}}, \citenamefont {Wambach},\ and\
  \citenamefont {Speth}}]{NISHIZAKI1988231}%
  \BibitemOpen
  \bibfield  {author} {\bibinfo {author} {\bibfnamefont {S.}~\bibnamefont
  {Nishizaki}}, \bibinfo {author} {\bibfnamefont {S.}~\bibnamefont
  {Dro\.{z}d\.{z}}}, \bibinfo {author} {\bibfnamefont {J.}~\bibnamefont
  {Wambach}}, \ and\ \bibinfo {author} {\bibfnamefont {J.}~\bibnamefont
  {Speth}},\ }\href {\doibase https://doi.org/10.1016/0370-2693(88)91425-6}
  {\bibfield  {journal} {\bibinfo  {journal} {Phys. Lett. B}\ }\textbf
  {\bibinfo {volume} {215}},\ \bibinfo {pages} {231 } (\bibinfo {year}
  {1988})}\BibitemShut {NoStop}%
\bibitem [{\citenamefont {Dinh~Dang}\ \emph {et~al.}(1997)\citenamefont
  {Dinh~Dang}, \citenamefont {Arima}, \citenamefont {Suzuki},\ and\
  \citenamefont {Yamaji}}]{NGUYENDINHDANG1997719}%
  \BibitemOpen
  \bibfield  {author} {\bibinfo {author} {\bibfnamefont {N.}~\bibnamefont
  {Dinh~Dang}}, \bibinfo {author} {\bibfnamefont {A.}~\bibnamefont {Arima}},
  \bibinfo {author} {\bibfnamefont {T.}~\bibnamefont {Suzuki}}, \ and\ \bibinfo
  {author} {\bibfnamefont {S.}~\bibnamefont {Yamaji}},\ }\href {\doibase
  https://doi.org/10.1016/S0375-9474(97)00172-3} {\bibfield  {journal}
  {\bibinfo  {journal} {Nucl. Phys.}\ }\textbf {\bibinfo {volume} {A621}},\
  \bibinfo {pages} {719 } (\bibinfo {year} {1997})}\BibitemShut {NoStop}%
\bibitem [{\citenamefont {Yannouleas}(1987)}]{Yannouleas1987}%
  \BibitemOpen
  \bibfield  {author} {\bibinfo {author} {\bibfnamefont {C.}~\bibnamefont
  {Yannouleas}},\ }\href@noop {} {\bibfield  {journal} {\bibinfo  {journal}
  {Phys. Rev. C}\ }\textbf {\bibinfo {volume} {35}},\ \bibinfo {pages} {1159}
  (\bibinfo {year} {1987})}\BibitemShut {NoStop}%
\bibitem [{\citenamefont {Rowe}(1970)}]{DJRowe}%
  \BibitemOpen
  \bibfield  {author} {\bibinfo {author} {\bibfnamefont {D.}~\bibnamefont
  {Rowe}},\ }\href@noop {} {\emph {\bibinfo {title} {Nuclear Collective
  Motion}}}\ (\bibinfo  {publisher} {Methuen, London},\ \bibinfo {year}
  {1970})\BibitemShut {NoStop}%
\bibitem [{\citenamefont {Ring}\ and\ \citenamefont
  {Schuck}(1980)}]{RingandSchuck}%
  \BibitemOpen
  \bibfield  {author} {\bibinfo {author} {\bibfnamefont {P.}~\bibnamefont
  {Ring}}\ and\ \bibinfo {author} {\bibfnamefont {P.}~\bibnamefont {Schuck}},\
  }\href@noop {} {\emph {\bibinfo {title} {The Nuclear Many-Body Problem}}}\
  (\bibinfo  {publisher} {Springer-Verlag, Berlin},\ \bibinfo {year}
  {1980})\BibitemShut {NoStop}%
\bibitem [{\citenamefont {Petrovich}\ and\ \citenamefont
  {Stanley}(1977)}]{Petrovich1977}%
  \BibitemOpen
  \bibfield  {author} {\bibinfo {author} {\bibfnamefont {F.}~\bibnamefont
  {Petrovich}}\ and\ \bibinfo {author} {\bibfnamefont {D.}~\bibnamefont
  {Stanley}},\ }\href {\doibase http://dx.doi.org/10.1016/0375-9474(77)90465-1}
  {\bibfield  {journal} {\bibinfo  {journal} {Nucl. Phys.}\ }\textbf {\bibinfo
  {volume} {A275}},\ \bibinfo {pages} {487 } (\bibinfo {year}
  {1977})}\BibitemShut {NoStop}%
\bibitem [{\citenamefont {Cook}\ \emph {et~al.}(1984)\citenamefont {Cook},
  \citenamefont {Kemper}, \citenamefont {Drumm}, \citenamefont {Fifield},
  \citenamefont {Hotchkis}, \citenamefont {Ophel},\ and\ \citenamefont
  {Woods}}]{Cook1984}%
  \BibitemOpen
  \bibfield  {author} {\bibinfo {author} {\bibfnamefont {J.}~\bibnamefont
  {Cook}}, \bibinfo {author} {\bibfnamefont {K.~W.}\ \bibnamefont {Kemper}},
  \bibinfo {author} {\bibfnamefont {P.~V.}\ \bibnamefont {Drumm}}, \bibinfo
  {author} {\bibfnamefont {L.~K.}\ \bibnamefont {Fifield}}, \bibinfo {author}
  {\bibfnamefont {M.~A.~C.}\ \bibnamefont {Hotchkis}}, \bibinfo {author}
  {\bibfnamefont {T.~R.}\ \bibnamefont {Ophel}}, \ and\ \bibinfo {author}
  {\bibfnamefont {C.~L.}\ \bibnamefont {Woods}},\ }\href {\doibase
  10.1103/PhysRevC.30.1538} {\bibfield  {journal} {\bibinfo  {journal} {Phys.
  Rev. C}\ }\textbf {\bibinfo {volume} {30}},\ \bibinfo {pages} {1538}
  (\bibinfo {year} {1984})}\BibitemShut {NoStop}%
\bibitem [{\citenamefont {Perey}\ and\ \citenamefont
  {Buck}(1962)}]{PEREY1962353}%
  \BibitemOpen
  \bibfield  {author} {\bibinfo {author} {\bibfnamefont {F.}~\bibnamefont
  {Perey}}\ and\ \bibinfo {author} {\bibfnamefont {B.}~\bibnamefont {Buck}},\
  }\href {\doibase http://dx.doi.org/10.1016/0029-5582(62)90345-0} {\bibfield
  {journal} {\bibinfo  {journal} {Nucl. Phys.}\ }\textbf {\bibinfo {volume}
  {32}},\ \bibinfo {pages} {353 } (\bibinfo {year} {1962})}\BibitemShut
  {NoStop}%
\bibitem [{\citenamefont {Van~Giai}\ and\ \citenamefont
  {Sagawa}(1981)}]{Phys.Lett.B106_379}%
  \BibitemOpen
  \bibfield  {author} {\bibinfo {author} {\bibfnamefont {N.}~\bibnamefont
  {Van~Giai}}\ and\ \bibinfo {author} {\bibfnamefont {H.}~\bibnamefont
  {Sagawa}},\ }\href@noop {} {\bibfield  {journal} {\bibinfo  {journal} {Phys.
  Lett. B}\ }\textbf {\bibinfo {volume} {106}},\ \bibinfo {pages} {379}
  (\bibinfo {year} {1981})}\BibitemShut {NoStop}%
\bibitem [{\citenamefont {Ohmura}\ \emph {et~al.}(1970)\citenamefont {Ohmura},
  \citenamefont {Imanishi}, \citenamefont {Ichimura},\ and\ \citenamefont
  {Kawai}}]{Ohmura1970}%
  \BibitemOpen
  \bibfield  {author} {\bibinfo {author} {\bibfnamefont {T.}~\bibnamefont
  {Ohmura}}, \bibinfo {author} {\bibfnamefont {B.}~\bibnamefont {Imanishi}},
  \bibinfo {author} {\bibfnamefont {M.}~\bibnamefont {Ichimura}}, \ and\
  \bibinfo {author} {\bibfnamefont {M.}~\bibnamefont {Kawai}},\ }\href@noop {}
  {\bibfield  {journal} {\bibinfo  {journal} {Prog. Theor. Phys.}\ }\textbf
  {\bibinfo {volume} {43}},\ \bibinfo {pages} {347} (\bibinfo {year}
  {1970})}\BibitemShut {NoStop}%
\bibitem [{\citenamefont {Koning}\ and\ \citenamefont
  {Delaroche}(2003)}]{KONING2003231}%
  \BibitemOpen
  \bibfield  {author} {\bibinfo {author} {\bibfnamefont {A.}~\bibnamefont
  {Koning}}\ and\ \bibinfo {author} {\bibfnamefont {J.}~\bibnamefont
  {Delaroche}},\ }\href {\doibase
  http://dx.doi.org/10.1016/S0375-9474(02)01321-0} {\bibfield  {journal}
  {\bibinfo  {journal} {Nucl. Phys.}\ }\textbf {\bibinfo {volume} {A713}},\
  \bibinfo {pages} {231 } (\bibinfo {year} {2003})}\BibitemShut {NoStop}%
\bibitem [{\citenamefont {Hama}\ \emph {et~al.}(1990)\citenamefont {Hama},
  \citenamefont {Clark}, \citenamefont {Cooper}, \citenamefont {Sherif},\ and\
  \citenamefont {Mercer}}]{HamaPhysRevC.41.2737}%
  \BibitemOpen
  \bibfield  {author} {\bibinfo {author} {\bibfnamefont {S.}~\bibnamefont
  {Hama}}, \bibinfo {author} {\bibfnamefont {B.~C.}\ \bibnamefont {Clark}},
  \bibinfo {author} {\bibfnamefont {E.~D.}\ \bibnamefont {Cooper}}, \bibinfo
  {author} {\bibfnamefont {H.~S.}\ \bibnamefont {Sherif}}, \ and\ \bibinfo
  {author} {\bibfnamefont {R.~L.}\ \bibnamefont {Mercer}},\ }\href {\doibase
  10.1103/PhysRevC.41.2737} {\bibfield  {journal} {\bibinfo  {journal} {Phys.
  Rev. C}\ }\textbf {\bibinfo {volume} {41}},\ \bibinfo {pages} {2737}
  (\bibinfo {year} {1990})}\BibitemShut {NoStop}%
\bibitem [{\citenamefont {Satchler}\ \emph {et~al.}(1964)\citenamefont
  {Satchler}, \citenamefont {Drisko},\ and\ \citenamefont
  {Bassel}}]{SatchlerPhysRev.136.B637}%
  \BibitemOpen
  \bibfield  {author} {\bibinfo {author} {\bibfnamefont {G.~R.}\ \bibnamefont
  {Satchler}}, \bibinfo {author} {\bibfnamefont {R.~M.}\ \bibnamefont
  {Drisko}}, \ and\ \bibinfo {author} {\bibfnamefont {R.~H.}\ \bibnamefont
  {Bassel}},\ }\href {\doibase 10.1103/PhysRev.136.B637} {\bibfield  {journal}
  {\bibinfo  {journal} {Phys. Rev.}\ }\textbf {\bibinfo {volume} {136}},\
  \bibinfo {pages} {B637} (\bibinfo {year} {1964})}\BibitemShut {NoStop}%
\bibitem [{\citenamefont {Burrows}(2006)}]{BURROWS20061747}%
  \BibitemOpen
  \bibfield  {author} {\bibinfo {author} {\bibfnamefont {T.}~\bibnamefont
  {Burrows}},\ }\href {\doibase http://dx.doi.org/10.1016/j.nds.2006.05.005}
  {\bibfield  {journal} {\bibinfo  {journal} {Nuclear Data Sheets}\ }\textbf
  {\bibinfo {volume} {107}},\ \bibinfo {pages} {1747 } (\bibinfo {year}
  {2006})}\BibitemShut {NoStop}%
\bibitem [{\citenamefont {Lane}(1962)}]{Lane1962676}%
  \BibitemOpen
  \bibfield  {author} {\bibinfo {author} {\bibfnamefont {A.~M.}\ \bibnamefont
  {Lane}},\ }\href {\doibase http://dx.doi.org/10.1016/0029-5582(62)90153-0}
  {\bibfield  {journal} {\bibinfo  {journal} {Nucl. Phys.}\ }\textbf {\bibinfo
  {volume} {35}},\ \bibinfo {pages} {676 } (\bibinfo {year}
  {1962})}\BibitemShut {NoStop}%
\bibitem [{\citenamefont {Doering}\ \emph {et~al.}(1975)\citenamefont
  {Doering}, \citenamefont {Patterson},\ and\ \citenamefont
  {Galonsky}}]{DoeringPhysRevC.12.378}%
  \BibitemOpen
  \bibfield  {author} {\bibinfo {author} {\bibfnamefont {R.~R.}\ \bibnamefont
  {Doering}}, \bibinfo {author} {\bibfnamefont {D.~M.}\ \bibnamefont
  {Patterson}}, \ and\ \bibinfo {author} {\bibfnamefont {A.}~\bibnamefont
  {Galonsky}},\ }\href {\doibase 10.1103/PhysRevC.12.378} {\bibfield  {journal}
  {\bibinfo  {journal} {Phys. Rev. C}\ }\textbf {\bibinfo {volume} {12}},\
  \bibinfo {pages} {378} (\bibinfo {year} {1975})}\BibitemShut {NoStop}%
\bibitem [{\citenamefont {Jon}\ \emph {et~al.}(2000)\citenamefont {Jon},
  \citenamefont {Orihara}, \citenamefont {Yun}, \citenamefont {Terakawa},
  \citenamefont {Itoh}, \citenamefont {Yamamoto}, \citenamefont {Suzuki},
  \citenamefont {Mizuno}, \citenamefont {Kamurai}, \citenamefont {Ishii},\ and\
  \citenamefont {Ohnuma}}]{JonPhysRevC.62.044609}%
  \BibitemOpen
  \bibfield  {author} {\bibinfo {author} {\bibfnamefont {G.~C.}\ \bibnamefont
  {Jon}}, \bibinfo {author} {\bibfnamefont {H.}~\bibnamefont {Orihara}},
  \bibinfo {author} {\bibfnamefont {C.~C.}\ \bibnamefont {Yun}}, \bibinfo
  {author} {\bibfnamefont {A.}~\bibnamefont {Terakawa}}, \bibinfo {author}
  {\bibfnamefont {K.}~\bibnamefont {Itoh}}, \bibinfo {author} {\bibfnamefont
  {A.}~\bibnamefont {Yamamoto}}, \bibinfo {author} {\bibfnamefont
  {H.}~\bibnamefont {Suzuki}}, \bibinfo {author} {\bibfnamefont
  {H.}~\bibnamefont {Mizuno}}, \bibinfo {author} {\bibfnamefont
  {G.}~\bibnamefont {Kamurai}}, \bibinfo {author} {\bibfnamefont
  {K.}~\bibnamefont {Ishii}}, \ and\ \bibinfo {author} {\bibfnamefont
  {H.}~\bibnamefont {Ohnuma}},\ }\href {\doibase 10.1103/PhysRevC.62.044609}
  {\bibfield  {journal} {\bibinfo  {journal} {Phys. Rev. C}\ }\textbf {\bibinfo
  {volume} {62}},\ \bibinfo {pages} {044609} (\bibinfo {year}
  {2000})}\BibitemShut {NoStop}%
\bibitem [{\citenamefont {Khoa}\ \emph {et~al.}(2007)\citenamefont {Khoa},
  \citenamefont {Than},\ and\ \citenamefont {Cuong}}]{KhoaPhysRevC.76.014603}%
  \BibitemOpen
  \bibfield  {author} {\bibinfo {author} {\bibfnamefont {D.~T.}\ \bibnamefont
  {Khoa}}, \bibinfo {author} {\bibfnamefont {H.~S.}\ \bibnamefont {Than}}, \
  and\ \bibinfo {author} {\bibfnamefont {D.~C.}\ \bibnamefont {Cuong}},\ }\href
  {\doibase 10.1103/PhysRevC.76.014603} {\bibfield  {journal} {\bibinfo
  {journal} {Phys. Rev. C}\ }\textbf {\bibinfo {volume} {76}},\ \bibinfo
  {pages} {014603} (\bibinfo {year} {2007})}\BibitemShut {NoStop}%
\bibitem [{\citenamefont {Gambacurta}\ \emph {et~al.}(2010)\citenamefont
  {Gambacurta}, \citenamefont {Grasso},\ and\ \citenamefont
  {Catara}}]{Phys.Rev.C81.054312}%
  \BibitemOpen
  \bibfield  {author} {\bibinfo {author} {\bibfnamefont {D.}~\bibnamefont
  {Gambacurta}}, \bibinfo {author} {\bibfnamefont {M.}~\bibnamefont {Grasso}},
  \ and\ \bibinfo {author} {\bibfnamefont {F.}~\bibnamefont {Catara}},\ }\href
  {\doibase 10.1103/PhysRevC.81.054312} {\bibfield  {journal} {\bibinfo
  {journal} {Phys. Rev. C}\ }\textbf {\bibinfo {volume} {81}},\ \bibinfo
  {pages} {054312} (\bibinfo {year} {2010})}\BibitemShut {NoStop}%
\bibitem [{\citenamefont {Franey}\ and\ \citenamefont
  {Love}(1985)}]{FLPhysRevC.31.488}%
  \BibitemOpen
  \bibfield  {author} {\bibinfo {author} {\bibfnamefont {M.~A.}\ \bibnamefont
  {Franey}}\ and\ \bibinfo {author} {\bibfnamefont {W.~G.}\ \bibnamefont
  {Love}},\ }\href {\doibase 10.1103/PhysRevC.31.488} {\bibfield  {journal}
  {\bibinfo  {journal} {Phys. Rev. C}\ }\textbf {\bibinfo {volume} {31}},\
  \bibinfo {pages} {488} (\bibinfo {year} {1985})}\BibitemShut {NoStop}%
\bibitem [{\citenamefont {Jeukenne}\ \emph {et~al.}(1977)\citenamefont
  {Jeukenne}, \citenamefont {Lejeune},\ and\ \citenamefont
  {Mahaux}}]{JeukennePhysRevC.16.80}%
  \BibitemOpen
  \bibfield  {author} {\bibinfo {author} {\bibfnamefont {J.-P.}\ \bibnamefont
  {Jeukenne}}, \bibinfo {author} {\bibfnamefont {A.}~\bibnamefont {Lejeune}}, \
  and\ \bibinfo {author} {\bibfnamefont {C.}~\bibnamefont {Mahaux}},\ }\href
  {\doibase 10.1103/PhysRevC.16.80} {\bibfield  {journal} {\bibinfo  {journal}
  {Phys. Rev. C}\ }\textbf {\bibinfo {volume} {16}},\ \bibinfo {pages} {80}
  (\bibinfo {year} {1977})}\BibitemShut {NoStop}%
\bibitem [{\citenamefont {Kerman}\ \emph {et~al.}(1959)\citenamefont {Kerman},
  \citenamefont {McManus},\ and\ \citenamefont {Thaler}}]{KERMAN1959551}%
  \BibitemOpen
  \bibfield  {author} {\bibinfo {author} {\bibfnamefont {A.}~\bibnamefont
  {Kerman}}, \bibinfo {author} {\bibfnamefont {H.}~\bibnamefont {McManus}}, \
  and\ \bibinfo {author} {\bibfnamefont {R.}~\bibnamefont {Thaler}},\ }\href
  {\doibase http://dx.doi.org/10.1016/0003-4916(59)90076-4} {\bibfield
  {journal} {\bibinfo  {journal} {Ann. Phys.}\ }\textbf {\bibinfo {volume}
  {8}},\ \bibinfo {pages} {551 } (\bibinfo {year} {1959})}\BibitemShut
  {NoStop}%
\bibitem [{\citenamefont {Bertsch}\ and\ \citenamefont
  {Esbensen}(1987)}]{BE0034-4885-50-6-001}%
  \BibitemOpen
  \bibfield  {author} {\bibinfo {author} {\bibfnamefont {G.~F.}\ \bibnamefont
  {Bertsch}}\ and\ \bibinfo {author} {\bibfnamefont {H.}~\bibnamefont
  {Esbensen}},\ }\href {http://stacks.iop.org/0034-4885/50/i=6/a=001}
  {\bibfield  {journal} {\bibinfo  {journal} {Rep. Prog. Phys.}\ }\textbf
  {\bibinfo {volume} {50}},\ \bibinfo {pages} {607} (\bibinfo {year}
  {1987})}\BibitemShut {NoStop}%
\end{thebibliography}%
%+++++++++++++++++++++++++++++++++++++++++++++++++++++++++++++++++++++

\end{document}